\documentclass[journal,10pt,draftclsnofoot,onecolumn]{IEEEtran}
\IEEEoverridecommandlockouts
\usepackage[noadjust]{cite}

\usepackage{amsmath,amssymb,amsfonts}
\usepackage{algorithmic}
\usepackage{graphicx}
\usepackage{textcomp}
\usepackage{xcolor}
\usepackage[all]{xy} 
\usepackage{bm}
\usepackage{multirow}

\def\BibTeX{{\rm B\kern-.05em{\sc i\kern-.025em b}\kern-.08em
    T\kern-.1667em\lower.7ex\hbox{E}\kern-.125emX}}

\newcommand{\red}[1]{\textup{{\color{red}#1}}}

\newcommand{\beq}{\begin{equation}}
\newcommand{\enq}{\end{equation}}
\newcommand{\bel}{\begin{lemma}}
\newcommand{\enl}{\end{lemma}}
\newcommand{\bet}{\begin{theorem}}
\newcommand{\ent}{\end{theorem}}

\newcommand*{\cC}{\Co}

\newcommand{\suppress}[1]{}

\mathchardef\mhyphen="2D

\def\sM{\mathsf{M}}
\def\sL{\mathsf{L}}
\def\sN{\mathsf{N}}

\makeatletter
\newcommand*{\rom}[1]{\expandafter\@slowromancap\romannumeral #1@}
\makeatother

\mathchardef\mhyphen="2D

\newtheorem{remark}{Remark}
\newtheorem{theorem}{Theorem}
\newtheorem{lemma}{Lemma}
\newtheorem{corollary}{Corollary}

\makeatletter

\newcommand{\Rmnum}[1]{\expandafter\@slowromancap\romannumeral #1@}
\makeatother

\newcommand{\FF}{\mathbb{F}}
\newcommand{\RR}{\mathbb{R}}
\newcommand{\Co}{C}

\graphicspath{{./fig/}}
\DeclareGraphicsExtensions{.pdf} 

\def\QED{\mbox{\rule[0pt]{1.5ex}{1.5ex}}}

 \newenvironment{proofof}[1]{\vspace*{5mm} \par \noindent
         \quad{\it Proof of #1:\hspace{2mm}}}{\QED
}
\def\Unif{\mathop{\rm Unif}}

\def\inte{\mathsf{inn}}

\def\Label#1{\label{#1}\ [\ \text{#1}\ ]\ }
\def\Label{\label}

\begin{document}

\title{Non-Adaptive Coding for Two-Way Wiretap Channel with or without Cost Constraints}

\author{%
Masahito~Hayashi,~\IEEEmembership{Fellow,~IEEE}, and~
Yanling Chen,~\IEEEmembership{Member,~IEEE} 
\thanks{
Masahito Hayashi is with 
School of Data Science, The Chinese University of Hong Kong, Shenzhen, Longgang District, Shenzhen, 518172, China,
International Quantum Academy (SIQA), Futian District, Shenzhen 518048, China,
and
Graduate School of Mathematics, Nagoya University, Chikusa-ku, Nagoya 464-8602, Japan.
(e-mail: hmasahito@cuhk.edu.cn, hayashi@iqasz.cn).}
\thanks{Yanling Chen is with Volkswagen Infotainment GmbH, Germany.
(e-mail: yanling.chen@volkswagen-infotainment.com).}
\thanks{MH is supported in part by the National Natural Science Foundation of China (Grant No. 62171212).}
}
\markboth{Non-Adaptive Coding for Two-Way Wiretap Channel with or without Cost Constraints}{}
\maketitle

\begin{abstract}
This paper studies the secrecy results for the two-way wiretap channel (TW-WC) with an external eavesdropper under a strong secrecy metric. Employing non-adaptive coding, we analyze the {\it information leakage} and the decoding error probability, and derive inner bounds on the secrecy capacity regions for the TW-WC under strong joint and individual secrecy constraints. For the TW-WC without cost constraint, both the secrecy and error exponents could be characterized by the {\it conditional R\'enyi mutual information} 
in a concise and compact form. And, some special cases secrecy capacity region and sum-rate capacity results are established, demonstrating that adaption is useless in some cases or the maximum sum-rate that could be achieved by non-adaptive coding. For the TW-WC with cost constraint, we consider the peak cost constraint and extend our secrecy results by using the constant composition codes. Accordingly, we characterize both the secrecy and error exponents by 
{\it a modification of R\'enyi mutual information}, which yields inner bounds on the secrecy capacity regions for the general discrete memoryless TW-WC with cost constraint. Our method works even when a pre-noisy processing is employed based on a conditional distribution in the encoder and can be easily extended to other multi-user communication scenarios.
\end{abstract}


\section{Introduction}
\subsection{Two-way channel}
The two-way communications channel (TWC) was first introduced by Shannon in a pioneering paper \cite{Shannon1961}, where he established an inner bound and an outer bound for the capacity region of a discrete memoryless TWC. 
18 year later, Dueck \cite{Dueck1979} provided an example showing that the capacity region of the TWC can exceed Shannon's inner bound. Thereafter, Schalkwijk \cite{Schalkwijk1983} used a constructive coding approach for the binary multiplier channel (BMC), attaining an achievable rate pair that is beyond Shannon's inner bound; whilst a more general coding scheme for TWC was proposed by Han in \cite{Han1984}. Til now the BMC still serves as an example of deterministic, binary, common output TWC where capacity in term of single-letter characterization is not yet known so far.

Why does Shannon's inner bound not coincide with Shannon's outer bound in general? The reason is that the inner bound is achieved using non-adaptive encoding, where the channel input at each user is constructed from the message to be transmitted only (and is independent of the past received signals). Clearly, this non-adaptive encoding makes the channel inputs of the two users independent. However, the outer bound allows the channel inputs of the two users dependent, where the dependence could be introduced by the adaption of channel input to the past received signals at each user.

Interestingly, there are certain special cases where adaptation is useless from a capacity perspective, making Shannon's inner bound coincide with Shannon's outer bound. 
These special cases include but not limited to:
\begin{itemize}
	\item the class of symmetric discrete memoryless TWCs \cite{Shannon1961};
	\item  the additive white Gaussian noise (AWGN) TWC \cite{Han1984};
	\item the additive exponential noise
	TWC \cite{Varshney2013};
	\item the finite-field additive
	noise TWC \cite{SAL2016};
	\item the class of injective semideterministic TWCs \cite{CVA2017}.
\end{itemize}
For these channels, the signals received in the past are not needed for the encoding to achieve any rate pair within the capacity region, rendering the TWC equivalent to two independent one-way channels. This feature is desirable in practice for a simpler engineering system design. 

\subsection{Two-way wiretap channel (preceding studies)} 
Information theoretic secrecy was first developed by Shannon
in \cite{Shannon1949}, where the perfect secrecy is achieved if the eavesdropper's observation $\mathbf{Z}$ provides no information on the transmitted message $M$ (i.e., the information leakage to the eavesdropper $I(M;\mathbf{Z})=0$). Wyner in \cite{Wyner1975} applied this concept to the one-way discrete memoryless channel, which is known as ``wiretap channel"; and generalized the ``perfect secrecy" to an 
``asymptotic perfect secrecy" by measuring the equivocation at the eavesdropper by the normalized conditional entropy of the message $M$ given the  eavesdropper's observation $\mathbf{Z}$. Wyner defined a secrecy capacity, up to which it is possible to achieve the asymptotic perfect secrecy in the manner that the information leakage rate to the eavesdropper is negligible as compared to the number of channel uses $I(M;\mathbf{Z})/H(M) \to 0.$ Such a secrecy measured by information leakage rate is often referred as ``weak secrecy", as it tolerates the fact that the eavesdropper might obtain a substantial amount of information in an absolute sense \cite{src:Csiszar1996,MH2006}, i.e.,  $I(M;\mathbf{Z}) \nrightarrow 0.$

The two-way wiretap channel (TW-WC) was first considered in \cite{TY2007} for both the Gaussian TW-WC (with details in \cite{TY2008}) and the binary additive TW-WC, where two users are communicating with each other in the presence of an external
eavesdropper. Based on the observation that the eavesdropper receives signals through a multi-access channel (MAC), the references \cite{TY2007, TY2008} showed that the controlled interference between transmitted signals (i.e., cooperative jamming) from the legitimate users could provide secrecy gains.  Inner bounds on the secrecy capacity regions (subject to weak secrecy constraint: $I(M_1, M_2;\mathbf{Z})/H(M_1, M_2) \to 0$)
for both channels were derived. Note that these achievable secret-rate regions were established by using non-adaptive coding, since the cooperative jamming strategy used therein did not exploit the advantage of the previous received signals for the channel input. 

Can the inner bounds of the capacity regions be improved by utilizing these previous received signals (i.e., adaptive coding)?  The impact of these signals was studied in \cite{HY2013}. It was demonstrated that adaptive coding could be highly beneficial for certain Gaussian TW-WCs. Moreover, \cite{GKYG2013} considered the general discrete memoryless TW-WC channel, and derived an inner bound of the capacity region under weak secrecy by combining cooperative jamming and a secret-key exchange mechanism (i.e., one user sacrifices part of its secret rate to transmit a key to the other user). Remarkably, for the Gaussian TW-WC, the region obtained in \cite{GKYG2013} was shown to be strictly larger than the regions given in \cite{TY2008}, demonstrating the effectiveness of adaptive coding (each user utilizing a key received previously from the other user to encrypt part of its message as described in \cite{GKYG2013}) in terms of enlarging the existing secret-rate region achieved by non-adaptive coding in \cite{TY2008}.

The above mentioned studies on TW-WCs focused on the weak secrecy constraint that is defined by $\allowbreak I(M_1, M_2;\mathbf{Z})\allowbreak/H(M_1, M_2) \to 0$ (which considers the leakage rate of $M_1, M_2$ jointly). Notably, there are also other secrecy measures for multi-user communications \cite{CKV18}.  Especially, for the TW-WCs, there are one-sided secrecy  \cite{QDT2017}  and individual secrecy \cite{QCHT2016}:
\begin{description}
	\item[(WO)] one-sided secrecy considering the leakage of $M_1$ or $M_2$ that is defined by $I(M_1;\mathbf{Z})/H(M_1) \to 0$ or $I(M_2;\mathbf{Z})/H(M_2) \to 0$;
	\item[(WI)] individual secrecy considering the leakage of $M_1$ and $M_2$ that is defined by $I(M_1;\mathbf{Z})/H(M_1) \to 0$ and $I(M_2;\mathbf{Z})/H(M_2) \to 0$.
\end{description}
 Inner bounds on the secrecy capacity regions for the TW-WCs under such one-sided/individual secrecy constraints are established in \cite{QDT2017} and \cite{QCHT2016}, respectively. 

Besides these studies on TW-WCs under weak secrecy, the reference \cite{PB2011} considered the general discrete memoryless TW-WC channel under a strong joint secrecy constraint, where the information leakage to the eavesdropper (rather than the leakage rate) is required to negligible (i.e., $I(M_1, M_2;\mathbf{Z}) \to 0$). The reference \cite{PB2011} not only provided the inner bounds on the secrecy capacity regions
by using cooperation jamming only (i.e., non-adaptive coding), but also recovered the achievable secret-rate region obtained in \cite{GKYG2013} by using cooperative jamming and key exchange (i.e., adaptive coding as in \cite{GKYG2013}). A noteworthy observation is that the region by adaptive coding therein did not improve the bound on the secrecy sum-rate that could be achieved by simply using the non-adaptive coding.  

\subsection{Two-way wiretap channel without cost constraint (our contributions)}
In this paper, we study the TW-WC (see Fig. \ref{fig: TW-WTC with an external eavesdropper} subject to a strong  secrecy criterion that can be
\begin{description}
	\item[(SJ)] joint secrecy defined by $I(M_1, M_2;Z^n)\to 0$; Or
	\item[(SI)] individual secrecy defined by $I(M_1;Z^n)\to 0$ and $I(M_2;Z^n)\to 0$. 
\end{description}
By definition, the joint secrecy implies the individual secrecy,   resulting in the first relationship as given in \eqref{eq: capacity relationship}. The one-sided secrecy case is not included here since the analysis is straightforward by following the approach to the case of individual secrecy.
Focusing on non-adaptive coding, we analyze the information leakage and the decoding error probability and derive inner bounds on the secrecy capacity regions for the TW-WC under strong joint and individual secrecy constraints. 
Our inner bound under the joint secrecy 
coincides with the inner bound established in \cite[Corollary 1]{PB2011}; while 
our inner bound under the individual secrecy is similar to what the reference \cite{QDT2017} obtained for weak secrecy. 
To this end, different approaches from \cite{PB2011} are taken to analyze the information leakage and the decoding error probability. Interestingly, both the secrecy and error exponents could be characterized by the 
{\it conditional R\'enyi mutual information} in a concise and compact form. 
Thus, the region is guaranteed with both information leakage and decoding error probability decreasing exponentially in the code length. More specifically, our technique for strong secrecy analysis reflects the channel resolvability for the $k$-transmitter MAC,
which was obtained in the reference \cite{HC2019} by extending the method by the references \cite{MH2006,HM};
while our error exponent could be regarded as a generalization of Gallager's error exponent \cite{Gallager68}. 

Taking a close look into the additive TW-WC over a finite field, we establish the following special case results:
\begin{itemize}
	\item under certain condition (that the eavesdropper observes a degraded version of the outputs at both legitimate users), the obtained inner bounds under the strong joint secrecy actually equals to 
	the joint secrecy capacity region (also for weak secrecy) if only non-adaptive coding is permitted.
	\item under certain condition (that given inputs, more noise is added to the eavesdropper's channel), 
	the obtained inner bounds under the individual strong secrecy 
	coincides with the capacity region of the channel, thus serving as also 
	the capacity region under the individual weak secrecy. 
	Since the inner bounds be achieved by non-adaptive codes, so adaption is useless in this case.
\end{itemize} 
Our work extends the existing results on additive TW-WC \cite{TY2007, GKYG2013} in several directions, such as: considering a more general model (not limiting to the binary case), considering other secrecy criteria (not necessarily joint secrecy), and developing special case converse results and so on. 

\subsection{Wiretap channel with cost constraint (preceding studies)}
When we focus on Gaussian channel, it is natural to impose a cost constraint
on the code.
In the case of wiretap channel, the sender uses a stochastic encoder. That is, to send a specific message, the sender chooses the transmitted input
randomly among several possible inputs. We consider two kinds of cost constraints. In the first case, the cost constraint is imposed to all possible inputs, which is called the peak constraint.
In the second case,  the cost constraint is imposed to the averaged inputs, which is called the average constraint.

Clearly, 
the capacity under the peak constraint is upper bounded 
by the capacity under the average constraint.
In the case of Gaussian wiretap channel, the reference \cite{Leung-Yan} derived the capacity under the average constraint
with the weak secrecy.
That is, the reference \cite{Leung-Yan} showed the direct and converse parts under this setting.
The essential tool of the converse part with the weak secrecy is the entropy power inequality \cite{Blachman}.

Later, the reference \cite{Han} derived an achievable rate
under the strong secrecy
with the peak constraint 
for a general discrete memoryless wiretap channel and a Poisson wiretap channel.
To make the desired code,  they employed random coding method by modifying
the distribution to generate codes by taking the cost constraint into account.
Also, the references \cite[Appendix D]{HM}, \cite{Tyagi} derived similar results,
and the reference \cite{Yang} studied this case with the second order asymptotics.
Adding the time sharing to the 
achievable rate obtained by \cite{Han},
the reference \cite{Sreekumar} derived 
the capacity under the strong secrecy
with the peak constraint 
for a general discrete memoryless wiretap channel.
It showed the existence of a channel, in which,
the achievable rate by \cite{Han} does not match 
the capacity under the cost constraint
while the achievable rate by \cite{Han} matches
the capacity under the cost constraint
for less noisy case \cite[Corollary 1]{Sreekumar}.
For the Gaussian wiretap channel,
the references \cite{Han} and \cite[Appendix D-C]{HM} showed the achievability of the rate obtained by 
the reference \cite{Leung-Yan} under the strong secrecy with the average constraint,
and the reference \cite{Favano} showed the same achievability 
under the strong secrecy with the peak constraint
when the constraint satisfies a certain condition.

However, it is not so easy to extend their method to our problem setting.
Instead of their method, we focus on constant composition codes
because the cost of each codeword in a constant composition code
always equals the average cost with respect to the distribution to generate the code.
Csis\'{a}r and K\"{o}rner \cite{CK} studied the channel coding with constant composition codes.
Also, the reference \cite{Merhav} studied the channel coding with the random coding on constant composition codes.
The reference \cite[Section XII]{HM} extended the coding with constant composition codes
by \cite{CK} to the wiretap channel.
 Recently, the reference \cite{ChengGao} studied quantum soft covering for constant composition codes,
which is equivalent to 
channel resolvability for constant composition codes.
Therefore, we can expect that the combination of the above two methods yields
another proof of the direct part of wiretap channel with cost constraint.
However, this simple combination does not work
even for a single-sender discrete memoryless wiretap channel
due to the following reason.
An encoder for wiretap channel
often requires the use of pre-noisy processing based on a conditional distribution in the encoder.
A constant composition code with the pre-noisy processing
does not satisfy the peak constraint, in general.
Therefore, when the wiretap channel is not degraded, 
we need a more advanced method for the peak constraint.

\subsection{Two-way wiretap channel with cost constraint (our contributions)}
Although several papers studied wiretap channel with peak and average cost constraints for various channel models, 
preceding papers \cite{TY2007, TY2008,HY2013,GKYG2013} studied two-way wiretap channel only under the average cost constraint and with the weak secrecy for the  Gaussian TW-WC.
In this paper, we introduce the peak cost constraint to the codes for a two-way wiretap channel,
and extend our results on the capacity regions for the TW-WC under strong joint and individual secrecy constraints (the conditions (SJ) and (SI))
by using the constant composition codes. 
Accordingly, 
based on the random coding on constant composition codes,
we characterize both the secrecy and error exponents by {\it a modification of R\'enyi mutual information}, which yields inner bounds of the above capacity regions.
It is worth mentioning that the obtained regions are valid for general discrete memoryless TW-WC which includes not only the channels with finite-alphabet inputs and outputs, e.g. the additive TW-WC over a finite field.
Our method works even when a pre-noisy processing is employed based on a conditional distribution
in the encoder.
Also, our method with constant composition codes directly works only with finite-alphabet inputs.
However, since the continuous-alphabet inputs can be approximated by finite-alphabet inputs,
a modification of our method also works with non-finite-alphabet inputs and outputs, e.g. Gaussion TW-WC.

In addition, similar to the reference \cite{Leung-Yan}, 
we also obtained an outer bound for the capacity region for Gaussion TW-WC with a cost constraint
by using the entropy power inequality \cite{Blachman}.
Combining the above results,
we identify the special case where our codes actually achieves the secrecy capacity of the sum-rates under the joint secrecy criteria (also for weak secrecy) if only non-adaptive coding is permitted. 

\subsection{Outline}
The rest of the paper is organized as follows. In Section \ref{sec:model}, we describe the system model of a two-way wiretap channel, give definitions of a non-adaptive or adaptive code (according to different setups at the encoder and decoder), different secrecy constraints (weak/strong, individual/joint) and present our main results on inner bounds for the respective secrecy capacity regions. 
In Section \ref{sec: Renyi Lemmas}, we provide the definitions of  {\it R\'enyi mutual information} and  {\it conditional R\'enyi mutual information} that are used to characterize the secrecy and error exponents. 
In Section \ref{sec: Proof}, we derive our inner bounds on the capacity regions for the TW-WC under strong joint/individual secrecy constraint using non-adaptive coding when no cost constraint is imposed. 
More specifically, we characterize both the secrecy and error exponents, which shed some lights on the exponential decreasing rates in the leaked information and the probability of decoding errors with respect to the number of channel uses while using the non-adaptive coding.
In Section \ref{sec: Proof2}, we extend our results in the previous section by introducing cost constraints to the non-adaptive codes. 
Using the constant composition codes, we show our inner bounds for the respective secrecy capacity regions of a TW-WC under a cost constraint,
and provide the secrecy and error exponents accordingly.
In Sections \ref{Sec: Additive TW-WC} and \ref{Sec: Gaussion TW-WC}, we take a close look into the additive TW-WC over a finite field and the Gaussian TW-WC, respectively. Secrecy capacity results are established for certain special cases. Also, these two sections derive outer bounds for the capacity regions for these two special cases.
Section \ref{Sec: Conclusion} concludes the paper by discussing possible extensions of this work.

\section{Channel model and strong secrecy result}\Label{sec:model}
\subsection{Formulation}
We consider that 
two legitimate users, User 1 and User 2 intend to exchange 
messages with each other in the presence of an external eavesdropper.
The two legitimate users and the external eavesdropper
are connected via 
a discrete memoryless two-way wiretap channel, 
which is characterized by $P_{Y_1 Y_2 Z|X_1 X_2}.$ 
In this model, User $i$ has the input symbol set ${\cal X}_i$ and 
the output symbol set ${\cal Y}_i$ for $i=1,2$.
The external eavesdropper has the output symbol set ${\cal Z}$.
The channel model is shown in Fig. \ref{fig: TW-WTC with an external eavesdropper}. 
When ${\cal X}_2$ and ${\cal Y}_1$ are singleton, 
the above model is a one-to-one wiretap channel.
Hence, the two-way wiretap channel model includes 
the one-to-one wiretap channel model.

\begin{figure}[htbp]
	\centering
	\includegraphics[width=.60\textwidth]{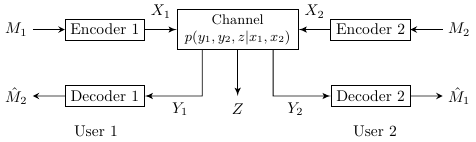}
	\caption{Two-Way wiretap channel with an external eavesdropper.} \Label{fig: TW-WTC with an external eavesdropper}
\end{figure}

In this paper, we use the following notations $\times$ and $\cdot$
for conditional distributions and distributions in the following way:
\begin{align}
P_{Z |X_1 X_2} \times P_{X_1|V_1} (z,x_1|v_1, x_2)&:=
P_{Z |X_1 X_2} (z|x_1,x_2)  P_{X_1|V_1} (x_1|v_1), \\
P_{Z |X_1 X_2} \cdot P_{X_1|V_1} (z|v_1,x_2)&:=
\sum_{x_1 \in {\cal X}_1}P_{Z |X_1 X_2} (z|x_1,x_2)  P_{X_1|V_1} (x_1|v_1), \\
P_{Z |X_1 X_2} \times P_{X_1|V_1} \times P_{X_2}(z, x_1,x_2|v_1)&:=
P_{Z |X_1 X_2} (z|x_1,x_2) P_{X_1|V_1} (x_1|v_1) P_{X_2}(x_2), \\
 P_{Z |X_1 X_2} \cdot P_{X_1|V_1} \times P_{X_2}(z, x_2|v_1)&:=
\sum_{x_1\in {\cal X}_1}P_{Z |X_1 X_2} (z|x_1,x_2) P_{X_1|V_1} (x_1|v_1) P_{X_2}(x_2) .
\end{align}

Their messages $M_i$ are assumed to be uniformly distributed over  the message sets
$\mathcal{M}_i=[1:e^{nR_i}]$ for $i=1,2$.
For their communication, they use the above channel $n$ times,
and 
we denote User $i$'s channel input and output 
by $X_i^n=(X_{i,1}, \ldots, X_{i,n}) \in {\cal X}_i^n$ and 
$Y_i^n=(Y_{i,1}, \ldots, Y_{i,n})\in {\cal Y}_i^n$, respectively.
Also, we denote the channel output at the eavesdropper by 
$Z^n=(Z_{1}, \ldots, Z_{n})\in {\cal Z}^n$. 

Consider the encoder of User $i$ for $i=1,2$. We consider two types of encoders for two users. The first type of encoder is a {\it non-adaptive} encoder $\phi_i$, which stochastically assigns the whole input $X_i^n$ based on the message $M_i$. The second type of encoder is an {\it adaptive} encoder $\phi_i$, which stochastically assigns the $t$-th input $X_{i,t}$ based on the message $M_i$ and 
the previous outputs $Y_{i,1},\ldots, Y_{i,t-1}$ for $t=1,\ldots,n$.

Consider the decoder of User $i$ for $i=1,2$.  We consider two types of decoders for two users.
In the 1st type decoder, 
the decoder $\psi_i$ of User $i$ is given as a map from (${\cal M}_i$, ${\cal X}_i^n$, ${\cal Y}_i^n$) to ${\cal M}_{i \oplus 1}.$
In the 2nd type decoder, 
the decoder $\psi_i$ of User $i$ is given as a map from 
(${\cal X}_i^n$, ${\cal Y}_i^n$) to ${\cal M}_{i \oplus 1}.$
Note that for two-way channels without any secrecy constraint, deterministic encoders are often used where user $i$ can recover information $X_i^n$ from $M_i$ and $Y_i^n.$
Hence, the 1st type decoder $\psi_i$ of User $i$ reduces to a map from (${\cal M}_i$, ${\cal Y}_i^n$) to 
${\cal M}_{i \oplus 1}$ as can be seen in previous works which include but not limited to \cite{Shannon1961, Varshney2013, CD2014, SAL2016, CVA2017, WSAL2019}.
However, for the TW-WC with secrecy constraint, stochastic encoders are often used where User $i$ cannot recover information $X_i^n$ from $M_i$ and $Y_i^n$ only. This is the reason why we consider the above two types of decoders.

When both encoders are non-adaptive encoders, we have the joint distribution as
\begin{align}
&P_{M_1 M_2 X_1^n X_2^n Y_1^n Y_2^n  Z^n}
(m_1, m_2, x_1^n, x_2^n, y_1^n, y_2^n, z^n)\nonumber \\
=&
P_{M_1}(m_1)
P_{M_2}(m_2) P_{X_1^n|M_1}(x_1^n|m_1)
P_{X_2^n|M_2}(x_2^n|m_2)P_{Y_1^n Y_2^n Z^n|X_1^n X_2^n}
(y_1^n, y_2^n,z^n|x_1^n, x_2^n),\Label{MVBY}
\end{align}
which implies the Markovian chains
\begin{align}
M_i-X_i^n-(M_{i \oplus 1},Y_i^n) \Label{MVI}
\end{align}
 for $i=1,2$.
Hence, 
knowing $M_i$ does not help 
the decoder $\psi_i$ of User $i$ (who holds (${\cal X}_i^n$, ${\cal Y}_i^n$)) gain any information on $M_{i \oplus 1}$ (here $\oplus$ denotes the modulo-2 addition). That is, it is sufficient to use the 2nd type decoder in this case.

Therefore, 
a $(e^{nR_1}, e^{nR_2}, n)$ adaptive (non-adaptive) secrecy code 
$\Co_n$ with the 1st (2nd) type decoder for the two-way wiretap channel consists of
$2$ message sets ($\mathcal{M}_1, \mathcal{M}_2$),
$2$ adaptive (non-adaptive) encoders ($\phi_1$, $\phi_2$), and
$2$ decoders ($\psi_1$, $\psi_2$) of the 1st (2nd) type.
When the encoder $\phi_i$ takes values only in a subset ${\cal S}_i \subset {\cal X}_i^n$ for $i=1,2$, we say that
a code $\Co_n$ is in ${\cal S}_1\times{\cal S}_2$.
We denote $(R_1,R_2)$ by $\kappa(C_n)$.

To evaluate the {\it reliability} of the transmission to the legitimate receiver, we consider the \emph{average probability of decoding error} at the legitimate receiver that is defined by
\begin{equation}\Label{eqn: Pe}
P_{e}^n(\Co_n)=\frac{1}{2^{n[R_1+R_2]}}\sum_{m_1, m_2}\Pr\left\{\bigcup\limits_{i\in \{1,2\}} \{m_i\neq \hat{m}_i\}\Bigg|\Co_n\right\}.
\end{equation}
 
To evaluate the {\it secrecy} of the transmission, we consider the strong secrecy that measures the amount of information (of $(M_1, M_2)$ jointly or of $M_1, M_2$ individually) leaked to the eavesdropper (through his/her observation $Z^n$).
We say that the rate pair $(R_{1},R_{2})$ is \emph{achievable under the strong joint secrecy constraint by adaptive (non-adaptive) codes with the 1st (2nd) type decoder}, if there exists a sequence of 
$(e^{nR_{1,n}}, e^{nR_{2,n}}, n)$ adaptive (non-adaptive) codes $\{\Co_n\}$ with the 1st (2nd) type decoder such that
$R_{i,n} \to R_i$ for $i=1,2$ and 
	\begin{align} 
	  P_{e}^n(\Co_n) &\leq \epsilon_n, \Label{eq:Reliability} \\
	  I(M_1, M_2;Z^n|\Co_n)	&\leq	\tau_n, \Label{eq:SSec}
\end{align}
with $\lim\limits_{n\to\infty} \epsilon_n = 0$ and $\lim\limits_{n\to\infty} \tau_n= 0.$ 
We say that the rate pair $(R_1,R_2)$ is \emph{achievable under the strong individual secrecy constraint by adaptive (non-adaptive) codes with the 1st (2nd) type decoder}, 
if there exists a sequence of $(e^{nR_{1,n}}, e^{nR_{2,n}}, n)$ adaptive (non-adaptive) codes $\{\Co_n\}$ with the 1st (2nd) type decoder such that
$R_{i,n} \to R_i$ for $i=1,2$ and 
	\begin{align} 
	  P_{e}^n(\Co_n) &\leq \epsilon_n, \Label{eq:Reliability2} \\
	  I(M_1;Z^n|\Co_n)	&\leq	\tau_n, \Label{eq:SSec2} \\
	  I( M_2;Z^n|\Co_n)	&\leq	\tau_n. \Label{eq:SSec3}
\end{align}

We define the achievability under the \emph{weak} joint secrecy constraint by adaptive (non-adaptive) codes
by replacing $\tau_n$ by $n \tau_n$ in \eqref{eq:SSec}.
In the same way, 
we define the achievability under the \emph{weak} 
individual secrecy constraint by adaptive (non-adaptive) codes
by replacing $\tau_n$ by $n \tau_n$ in \eqref{eq:SSec2} and \eqref{eq:SSec3}.

\subsection{Capacity region (without cost constraint)}
We have the following types of capacity regions.
\begin{align}
{\cal C}_{J,S(W), A(N),1(2)}&:=
\left\{ (R_1,R_2) \left| 
\begin{array}{l}
(R_1,R_2) \hbox{ is achievable under the strong (weak) joint secrecy}\\
\hbox{constraint by adaptive (non-adaptive) codes with the 1st (2nd) type decoder}
\end{array} 
\right.\right\},\\
{\cal C}_{I,S(W), A(N),1(2)}&:=
\left\{ (R_1,R_2) \left| 
\begin{array}{l}
(R_1,R_2) \hbox{ is achievable under the strong (weak) individual secrecy}\\
\hbox{constraint by adaptive (non-adaptive) codes with the 1st (2nd) type decoder}
\end{array} 
\right.\right\}.
\end{align}
We have the following relations
\begin{align}
{\cal C}_{J,b, c,d} \subset {\cal C}_{I,b, c,d}, ~
{\cal C}_{a,S, c,d} \subset {\cal C}_{a,W, c,d}, ~
{\cal C}_{a,b, N,d} \subset {\cal C}_{a,b,A,d},~
{\cal C}_{a,b, c,2} \subset {\cal C}_{a,b,c,1} \label{eq: capacity relationship}
\end{align}
for $a\in\{I,J\}$, $b\in \{S,W\}$, $c\in\{N,A\}$ and $d\in\{1,2\}.$
In addition, the Markovian chain \eqref{MVI} implies the relation
\begin{align}
{\cal C}_{a,b, N,2} = {\cal C}_{a,b,N,1}.
\end{align}

In wiretap channel coding, we often employ such a pre-noisy processing given by 
$P_{X_i|V_i}$ for $i=1,2$.
That is, the transmission rate of 
wiretap channel can be improved when
the encoder applies the pre-noisy processing given by a certain conditional distribution. 
Since we cannot deny this possibility in a two-way wiretap channel,  
we employ such pre-noisy processings to derive inner bounds of the above capacity regions.

Considering the strong joint secrecy criteria \eqref{eq:SSec}, we define the region
\begin{align}
{\cal R}_{J,N}(P_{V_1 V_2 X_1 X_2}):= 
\left\{(R_1,R_2) \in \mathbb{R}_+^2
\left|	
\begin{array}{l}
R_1 \leq I(Y_2;V_1|X_2)-I(Z;V_1), \\
R_2 \leq I(Y_1;V_2|X_1)-I(Z;V_2), \\
R_1+R_2  \leq  I(Y_2;V_1|X_2) + I(Y_1;V_2|X_1) - I(V_1, V_2;Z) 
\end{array}
	\right.\right\}
\Label{BPW1}
\end{align}
for a joint distribution $P_{V_1 V_2 X_1 X_2}$, where
$\mathbb{R}_+$ is the set of positive real numbers.\par
We also define ${\cal Q}:=\{\prod\limits_{i\in \{1,2\}}P_{V_i}(v_i)P_{X_i|V_i}(x_i|v_i)\}$.
Then, we define the region 
\begin{align}
{\cal R}_{J,N}:=& cl. \bigcup_{P_{V_1 V_2 X_1 X_2} \in {\cal Q}}{\cal R}_{J, N}(P_{V_1 V_2 X_1 X_2}),
\end{align}
where $cl.$ expresses the closure of the convex hull.

Considering the strong individual secrecy criteria \eqref{eq:SSec2} and \eqref{eq:SSec3}, we define the region
\begin{align}
{\cal R}_{I, N}(P_{V_1 V_2 X_1 X_2}):= 
\left\{(R_1,R_2)\in\mathbb{R}_+^2
\left|
\begin{array}{l}
R_1 \leq I(Y_2;V_1|X_2)-I(Z;V_1),\\
R_2 \leq I(Y_1;V_2|X_1)-I(Z;V_2),\\
\max\{R_1, R_2\} \leq I(Y_1;V_2|X_1)+I(Y_2;V_1|X_2)-I(V_1, V_2;Z)
\end{array}
\right.\right\}\Label{BPW2}
\end{align}
for a joint distribution $P_{V_1 V_2 X_1 X_2}.$
Then, we define the region 
\begin{align}
{\cal R}_{I,N}:= &cl. \bigcup_{P_{V_1 V_2 X_1 X_2}\in {\cal Q}}{\cal R}_{I,N}(P_{V_1 V_2 X_1 X_2}).
\end{align}

We have the following theorem for the achievable secrecy rate region of the two-way wiretap channel.

\begin{theorem}\Label{Cor: S-MAC-KS} 
We have the following relations.
\begin{align}
{\cal R}_{J,N} \subset {\cal C}_{J,S,N,2},\quad &
{\cal R}_{I,N} \subset {\cal C}_{I,S,N,2}.\Label{NZT}
\end{align}
\end{theorem} 

The reference \cite[Corollary 1]{PB2011}
also showed the same relation as the first relation in \eqref{NZT}.

\begin{remark}
In general, it is a difficult task to establish converse results for the general TW-WC, undoubtably more difficult than the task to derive converse results for the TWC (as TWC can be considered as a special case of TW-WC by taking $Z=\emptyset$). As a well-known fact, it still remains an open problem to establish the capacity region for the general TWC since TWC was first introduced by Shannon in \cite{Shannon1961}. 

Due to the fact that a TW-WC is a TWC under additional secrecy constraints,  any outer bound on the capacity region for a general TWC, e.g. Shannon's outer bound \cite{Shannon1961}; or capacity region established for a certain TWC, e.g. capacity region established for the finite field additive TWC \cite{SAL2016}, could serve as a trial outer bound on the secrecy capacity regions for the corresponding TW-WC as well. However, the derivation of a tight outer bound on the secrecy capacity regions is generally difficult because of the following reasons.  
	\begin{itemize}
		\item An adaptive encoder has many random variables to adaptively improve our code. In general, it is difficult to reduce the analysis into a single-letterized region. 
		\item We could derive an outer bound only when the channel is additive and the security criterion is individual, which is stated as \eqref{HH4-7B} in Theorem \ref{add-C}. It is unclear whether an adaptive encoder outperforms our obtained inner bounds in the general case.
		\item Note that a special case individual secrecy capacity region is established in Corollary \ref{add-C2} for the finite field additive TWC, where the individual secrecy capacity region for the TW-WC actually coincides with the capacity region for the TWC.
	\end{itemize}
Analysis is even still difficult for non-adaptive codes due to the following reasons.
\begin{itemize}
	\item Wiretap codes employ stochastic encoders. In the one-to-one channel, the method by Cisz\'{a}r and K\"{o}rner \cite[Theorem 17.11]{CK} converts a joint distribution on $X^n$ and $V^n$ on $n$ channel uses to a joint distribution on $X$ and $V$ on a single channel use.
	However, in the two-way channel case, we have a joint distribution $P_{X_1^n V_1^n} \times P_{X_2^n V_2^n}$ on $n$ channel uses. In this joint distribution, $X_1^n,V_1^n$ and $X_2^n,V_2^n$ are independent of each other. When we apply the same method as Cisz\'{a}r and K\"{o}rner, $X_1,V_1$ and $X_2,V_2$ might not be independent of each other under the converted joint distribution $P_{X_1 V_1 X_2 V_2}$ in general. 
	\item Fortunately, we could resolve this problem only for additive channels and Gaussian additive channels with certain conditions (by exploiting the channel characteristics instead of using the  Cisz\'{a}r-K\"{o}rner method). See Theorem \ref{add-C} and Theorem \ref{G-C}.
\end{itemize}
\end{remark}

\subsection{Capacity region (with cost constraint)}
Next, we introduce cost constraint for our codes.
We choose cost functions $g_1$ and $g_2$ on ${\cal X}_1$ and 
${\cal X}_2$, respectively.
Then, given two real numbers $c_1$ and $c_2$,
we impose the following cost constraint for a $(e^{nR_1}, e^{nR_2}, n)$
code $C_n=(\phi_1,\phi_2,\psi_1,\psi_2)$.
We define the subset 
${\cal X}_{i}^{n}(c_i)$
of ${\cal X}_{i}^{n}$ 
as
\begin{align}
{\cal X}_{i}^{n}(c_i):= 
\Bigg\{ (x_{i,1}, \ldots,x_{i,n})\in {\cal X}_{i}^{n}\Bigg|
\frac{1}{n} \sum_{j=1}^n  g_i(x_{i,j}) \le c_i
\Bigg\}
\end{align}
for $i=1,2$.
We say that the encoder satisfies the peak cost constraint with $c_1,c_2$ when
the input $X_i^n$ of the encoder $\phi_i$ takes values in the subset ${\cal X}_{i}^{n}(c_i)$ for $i=1,2$.
We say that the encoder satisfies the average cost constraint with $c_1,c_2$ when
the relation $\frac{1}{n}( \mathbb{E}  \sum_{j=1}^n  g_i(X_{i,j}) )\le c_i$
holds for $i=1,2$.

Then, we have the following types of capacity regions.
\begin{align}
{\cal C}_{J,S(W), A(N),1(2)}^{P(A),c_1,c_2}&:=
\left\{ (R_1,R_2) \left| 
\begin{array}{l}
(R_1,R_2) \hbox{ is achievable under the strong (weak) joint secrecy}\\
\hbox{constraint by adaptive (non-adaptive) codes with the 1st (2nd) type decoder}\\
\hbox{and with peak (average) cost constraint with }c_1,c_2.
\end{array} 
\right.\right\},\\
{\cal C}_{I,S(W), A(N),1(2)}^{P(A),c_1,c_2}&:=
\left\{ (R_1,R_2) \left| 
\begin{array}{l}
(R_1,R_2) \hbox{ is achievable under the strong (weak) individual secrecy}\\
\hbox{constraint by adaptive (non-adaptive) codes with the 1st (2nd) type decoder}\\
\hbox{and with peak (average) cost constraint with }c_1,c_2.
\end{array} 
\right.\right\}.
\end{align}
We have five indexes in total for characterizing the capacity regions as shown in Table \ref{Summary}. That is, we have $2^5=32$ capacity regions.

\begin{table}[t]
\caption{Summary of problem settings}
\label{Summary}
\begin{center}
\begin{tabular}{|l|c|c|c|c|c|}
\hline
Indexes &\multicolumn{2}{c|}{Secrecy criterion}   & Encoder  & Decoder & Cost constraint  \\
\hline 
Abbreviations & $J(I)$ & $S(W)$ & $A(N)$ & $1(2)$ & $P(A)$\\
\hline
\multirow{4}{*}{Definitions} & $J:$ Joint secrecy & $S:$ Strong secrecy & $A:$ Adaptive  & $1:$
1st type decoder & $P:$ Maximum of cost  \\
 & is considered. &  is imposed. & encoder is used.& 
is used.  &  is constrained. \\
\cline{2-6} 
 & $I:$ Individual  secrecy & $W:$  Weak secrecy & $N:$ Non-adaptive  & 
$2:$ 2nd type decoder & $A:$ Average of cost  \\
 & is considered. &  is imposed. & encoder is used.& 
is used.  &  is constrained. \\
\hline
\end{tabular}
\begin{flushleft}
In total, we have $2^5=32$ problem settings.
\end{flushleft}
\end{center}
\end{table}

We have the relation
\begin{align}
{\cal C}_{a,b,c,d}^{P,c_1,c_2}
\subset 
{\cal C}_{a,b, c,d}^{A,c_1,c_2},
\end{align}
for $a\in\{I,J\}$, $b\in \{S,W\}$, $c\in\{N,A\}$ and $d\in\{1,2\}.$

\begin{lemma}\Label{LL1}
For $a\in\{I,J\}$, $b\in \{S,W\}$, $c\in\{N,A\}$ and $d\in\{1,2\},$ the regions 
${\cal C}_{a,b,c,d}^{P,c_1,c_2}$ and ${\cal C}_{a,b, c,d}^{A,c_1,c_2}$
are closed sets.
\end{lemma}
\begin{IEEEproof}
See Appendix \ref{App: proof of Lemma LL1} for a detailed proof.
\end{IEEEproof}

For a joint distribution $P_{V_1 V_2 X_1 X_2}$, we define the condition:
\begin{align}
\sum_{x_1 \in {\cal X}_1}
P_{X_1 }(x_1) g_1(x_1) \le c_1
,\quad
\sum_{x_2 \in {\cal X}_2} 
P_{X_2}(x_2)g_2(x_2) \le c_2.\Label{NMZ}
\end{align}

We define the set ${\cal Q}^{c_1,c_2}:=
\Big\{\prod\limits_{i\in \{1,2\}}P_{V_i}(v_i)P_{X_i|V_i}(x_i|v_i)\Big|
$ the condition \eqref{NMZ} holds $\Big\}$.
Then, we define the regions 
\begin{align}
{\cal R}_{J,N}^{c_1,c_2}
:=&
cl. \bigcup_{P_{V_1 V_2 X_1 X_2} \in {\cal Q}^{c_1,c_2}}{\cal R}_{J,N}(P_{V_1 V_2 X_1 X_2}),
\\
{\cal R}_{I,N}^{c_1,c_2}
:=&
cl. \bigcup_{P_{V_1 V_2 X_1 X_2} \in {\cal Q}^{c_1,c_2}}{\cal R}_{I,N}(P_{V_1 V_2 X_1 X_2}).
\end{align}

As a generalization of Theorem \ref{Cor: S-MAC-KS},
we have the following theorem for the achievable secrecy rate region of the two-way wiretap channel with cost constraint.

\begin{theorem}\Label{Cor: C-MAC-KS} 
We have the following relations.
\begin{align}
{\cal R}_{J,N}^{c_1,c_2} \subset {\cal C}_{J,S,N,2}^{P,c_1,c_2},\quad
{\cal R}_{I,N}^{c_1,c_2} \subset {\cal C}_{I,S,N,2}^{P,c_1,c_2}.\Label{NZT2}
\end{align}
\end{theorem} 

Further, we extend the regions 
${\cal R}_{I,N}^{c_1,c_2}$
and
${\cal R}_{J,N}^{c_1,c_2}$ as
\begin{align}
\hat{{\cal R}}_{I(J),N}^{c_1,c_2}
:=\bigcup_{P, \vec{c}_1,\vec{c}_2 } 
\Big\{ \Big( \sum_{j}Q(j) R_1^j,\sum_{j}Q(j) R_2^j \Big)
\Big|(R_1^j, R_2^j)\in {\cal R}_{I(J),N}^{c_1^j,c_2^j} \Big\},
\end{align}
where $P, \vec{c}_1,\vec{c}_2 $ are chosen as follows.
$\vec{c}_i$ is a vector whose components take values in the interval between 
the infimum of the range of $g_i$ and the supremun of the range of $g_i$.
The numbers of components of $\vec{c}_1$ and $\vec{c}_2$
are arbitrarily chosen as the same number $n$.
$Q$ is a distribution on $\{1, \ldots, n\}$ to satisfy 
$\sum\limits_{j=1}^nQ(j) c_1^j \le c_1$
and $\sum\limits_{j=1}^nQ(j) c_2^j \le c_2$.
Considering time sharing in \eqref{NZT2}, we obtain the following relation.
\begin{align}
\hat{{\cal R}}_{J,N}^{c_1,c_2} \subset {\cal C}_{J,S,N,2}^{P,c_1,c_2},\quad
\hat{{\cal R}}_{I,N}^{c_1,c_2} \subset {\cal C}_{I,S,N,2}^{P,c_1,c_2}.\Label{NZT2B}
\end{align}
Here, we stress that the extension \eqref{NZT2B} is essential
because, 
as pointed in the reference \cite{Sreekumar},
there exists a 
one-to-one wiretap channel such that
the inner bound (with time-sharing) in \eqref{NZT2B} is strictly larger than 
the inner bound (without time-sharing) in \eqref{NZT2}. 

\section{Proof of Theorem \ref{Cor: S-MAC-KS}}\Label{sec: Proof}
To prove Theorem \ref{Cor: S-MAC-KS}, we need to find a sequence of non-adaptive codes ${\mathcal{C}_n}$ with the 2nd type decoder such that
\begin{itemize}
	\item the codes have rates pair converge to $(R_1, R_2)$ within the region ${\cal R}_{J,N}$ and ${\cal R}_{I,N}$ for joint secrecy and individual secrecy, respectively;
	\item the decoding error is bounded by $\epsilon_n$ with $\lim\limits_{n\to \infty} \epsilon_n =0;$
	\item the information leakage is bounded by $\tau_n$ with $\lim\limits_{n\to \infty} \tau_n =0.$
\end{itemize}
To this end, we propose a coding scheme and evaluate the decoding error (upper bounded by $\epsilon_n$) and respective information leakage (upper bounded by $\tau_n$) for joint secrecy and individual security. Especially, we show that both $\epsilon_n$ and $\tau_n$ are decreasing exponentially with respect to the code length $n$ (thus establishing the strong secrecy). As a preparation, in Sec. \ref{sec: Renyi Lemmas} we first give some definitions on the  (conditional) R\'enyi mutual information, which are used to quantize $\epsilon_n$ and $\tau_n$. The single-shot analysis is provided in Sec. \ref{sec: Single shot analysis}, laying foundations for the analysis of the coding scheme proposed in Sec. \ref{sec: Exponential evaluation}. The reliability and strong secrecy analysis of the coding scheme are provided in Sec. \ref{S4-B} and Sec. \ref{S4-C}, respectively. Sec. \ref{sec: Deriving the achievable region} derives the desired region in Theorem \ref{Cor: S-MAC-KS}.   

\subsection{Conditional R\'enyi mutual information}\Label{sec: Renyi Lemmas}

We recall the {\it R\'enyi relative entropy} as
\begin{align}\Label{eqn: Renyi relative entropy}
D_{1+s}(P \| Q):=\frac{1}{s}\log  \sum_{x}P(x)^{1+s} Q(x)^{-s}.
\end{align}
Note that $D_{1+s}(P \| Q)$ is nondecreasing w.r.t. $s$ for $s>0$ and $\lim\limits_{s\to 0}D_{1+s}(P \| Q)=D(P \| Q),$ i.e., the relative entropy.

We define the {\it R\'enyi mutual information} as
\begin{align}
 I_{1+s}^{\downarrow} ( Z ; X) 
:=
 D_{1+s} (P_{Z X}
\| P_Z \times P_{X}   ).\Label{Ne1}
\end{align}
and
the {\it conditional R\'enyi mutual information} as
\begin{align}
e^{-s I_{\frac{1}{1+s}}^{\uparrow} ( Z ; X|Y) }
:=
&
\sum_y P_Y(y)
e^{-s \min\limits_{Q_{Z|Y=y}}D_{\frac{1}{1+s}} (P_{Z X|Y=y} \|  Q_{Z|Y=y} \times P_{X|Y=y}   )}.
\end{align}
The minimizer $Q_{Z|Y}^*$ is given as
\begin{equation}
	Q_{Z|Y}^*(z|y)= \frac{\sum\limits_{ x} P_{X|Y}(x|y)
		P_{Z| X Y}(z|x,y)^{\frac{1}{1+s}}}{\sum\limits_{z'} \sum\limits_{ x'} P_{X|Y}(x'|y)
		P_{Z| X Y}(z'|x',y)^{\frac{1}{1+s}}},
\end{equation}
and $I_{\frac{1}{1+s}}^{\uparrow} ( Z ; X| Y )$
is written in the following form \cite[Eq. (54)]{TH};
\begin{align}
e^{-s I_{\frac{1}{1+s}}^{\uparrow} ( Z ; X| Y ) }
=
&\sum_{ y} P_Y(y)
\sum_{z} P_{Z|Y}(z|y)
\Big(
\sum_{ x} P_{X|Y}(x|y)^{\frac{s}{1+s}}
 P_{X| Z Y}(x|z,y)^{\frac{1}{1+s}}
\Big)^{1+s}  \nonumber\\
=
&\sum_{ y} P_Y(y)
\sum_{z} 
\Big(
\sum_{ x} P_{X|Y}(x|y)
 \Big(\frac{P_{ZX| Y}(z,x|y)}{ P_{X| Y}(x|y)}\Big)^{\frac{1}{1+s}}
\Big)^{1+s}  \nonumber\\
=
&\sum_{ y} P_Y(y)
\sum_{z} 
\Big(
\sum_{ x} P_{X|Y}(x|y)
 P_{Z| X Y}(z|x,y)^{\frac{1}{1+s}}
\Big)^{1+s} \Label{Ne3} .
\end{align}
When $X$ and $Y$ are independent of each other, 
it is simplified as
\begin{align}
e^{-s I_{\frac{1}{1+s}}^{\uparrow} ( Z ; X| Y ) }
=
&\sum_{ y} P_Y(y)
\sum_{z} 
\Big(
\sum_{ x} P_{X}(x)
 P_{Z| X Y}(z|x,y)^{\frac{1}{1+s}}
\Big)^{1+s}  \nonumber \\
=
&\sum_{ y,z} 
\Big(
\sum_{ x} P_{X}(x)
(P_Y(y) P_{Z| X Y}(z|x,y))^{\frac{1}{1+s}}
\Big)^{1+s} 
=e^{-s I_{\frac{1}{1+s}}^{\uparrow} ( Z Y; X ) }
\Label{Ne3Y} \\
=&
e^{-s \min\limits_{Q_{ZY}}D_{\frac{1}{1+s}} (P_{Z YX} \|  Q_{ZY} \times P_{X}   )}.
\Label{Ne3G} 
\end{align}
In addition, the minimizer $Q_{ZY}^*$ in \eqref{Ne3G} is given as
\begin{equation}
	Q_{ZY}^*(z,y)=
	\frac{\sum\limits_{ x} P_{X}(x)
		P_{ZY| X }(z,y|x)^{\frac{1}{1+s}}}{\sum\limits_{z',y'} \sum\limits_{ x'} P_{X}(x')
		P_{ZY| X }(z',y'|x')^{\frac{1}{1+s}}}.
\end{equation}
When we need to identify the joint distribution of $ZY$ and $X$, we denote
 it as
 $I_{\frac{1}{1+s}}^{\uparrow} ( Z Y; X )[P_{ZYX}] $.

Also, we have the inequality \cite[Lemma 1]{H2011}
\begin{align}
e^{s I_{1+s}^{\downarrow} ( Z ; X) }
\le
e^{s I_{\frac{1}{1-s}}^{\uparrow} ( Z ; X) }.\Label{MMK}
\end{align}
When $P_{Z|X} \le c \cdot Q_{Z|X}$ with a constant $c$, i.e., 
$P_{Z|X}(z|x) \le c \cdot Q_{Z|X}(z|x)$, the expression \eqref{Ne3Y} implies
\begin{align}
e^{s I_{\frac{1}{1-s}}^{\uparrow} ( Z ; X) [P_{Z|X}\times P_X]}
\le
c \cdot e^{s I_{\frac{1}{1-s}}^{\uparrow} ( Z ; X) [Q_{Z|X}\times P_X]}.\Label{MVK}
\end{align}
This quantity coincides with the Gallager function \cite{Gallager68}.

As another version of R\'{e}nyi mutual information, 
the paper \cite[(6)]{ChengGao} introduced a modification 
$\breve{I}_{\frac{1}{1+s}}^{\uparrow} ( Z ; X)$
of $I_{\frac{1}{1+s}}^{\uparrow} ( Z ; X)$ as
\begin{align}
\breve{I}_{\frac{1}{1+s}}^{\uparrow} ( Z ; X)
:=&
\min_{Q_{Z}}
\sum_x P_{X}(x)
D_{\frac{1}{1+s}} (P_{Z| X=x} \|  Q_{Z}  ).
\end{align}
Then, 
$\breve{I}_{\frac{1}{1+s}}^{\uparrow} ( Z ; X|Y)$ is defined as
\begin{align}
e^{-s \breve{I}_{\frac{1}{1+s}}^{\uparrow} ( Z ; X|Y)}
:=&
\sum_y P_Y(y) e^{-s \min_{Q_{Z|Y=y}}
\sum_x P_{X|Y}(x|y)
D_{\frac{1}{1+s}} (P_{Z| X=x,Y=y} \|  Q_{Z|Y=y}  )}.
\end{align}
When $X$ and $Y$ are independent of each other, 
as a generalization of above quantity,
we have 
$\breve{I}_{\frac{1}{1+s}}^{\uparrow} ( Z ; X|Y)
:=\breve{I}_{\frac{1}{1+s}}^{\uparrow} ( ZY ; X)$.

\subsection{Single shot analysis}\label{sec: Single shot analysis}
To show Theorem \ref{Cor: S-MAC-KS},
we prepare single shot analysis.
We have the following resolvability lemma.
\begin{lemma}\Label{L2}
Let $P_{Z|X_1 X_2}$ be a MAC with two input systems.
Assume that $P_{X_1 X_2}=P_{X_1} \times P_{X_2}$.
Let $X_{i,1},\ldots, X_{i,\sL_i}$ be the random variables that are independently generated subject to $P_{X_i},$ for $i=1,2$.
Here, we denote this code selection by ${\cC}$ and define
the distribution $P_{Z| \cC}$ for the variable $Z$ dependently of 
the code selection ${\cC}$ as
$
P_{Z| \cC}(z):=
\frac{1}{\sL_1 \sL_2} 
\sum\limits_{j_1,j_2} P_{Z| X_1=X_{1,j_1}, X_2=X_{2,j_2}  }(z).
$
Then, for $s \in [0,1]$, we have $ 
D(  P_{Z| \cC}\| P_{Z}) 
\le  
D_{1+s}(  P_{Z| \cC}\| P_{Z})$ and
\begin{align*}
s \mathbb{E}_{\cC} 
\left[D_{1+s}(  P_{Z| \cC}\| P_{Z}) \right]
\le 
\sum_{\mathcal{S}\neq \emptyset, \mathcal{S}\subset \{1,2\}}
\frac{1}{ \prod_{i \in \mathcal{S}} \sL_i^s} 
e^{s I^{\downarrow}_{1+s} ( Z ; X_{\mathcal{S}}) }
\end{align*}
for $s \in [0,1]$.
\end{lemma}
\begin{IEEEproof}
See Appendix \ref{A-E} for a detailed proof.
\end{IEEEproof}

\begin{remark}
Note that Lemma \ref{L2} is a generalization of a special case of \cite[Theorem 14]{HM}. (Also, the paper \cite{YT2019} uses a method similar to \cite[Theorem 14]{HM}.) 
Its extension to $k$-transmitter MAC is given in
the reference \cite[Lemma 1]{HC2019}.
Although the reference \cite{Sultana} discussed a similar type of 
channel resolvability, 
it assumes symmetric structure in the same way as \cite[Theorem 17]{HM}. Besides, Lemma  \ref{L2} can be regarded as an extension of the resolvability for the 2-transmitter MAC \cite{src:Steinberg98} by tying the strong secrecy to channel resolvability and further characterizing the secrecy exponent of the coding mechanism. 
\end{remark}

For reliability analysis, we have the following lemma 
as a generalization of Gallager's bound \cite{Gallager68} for two-transmitter MAC.
\begin{lemma}\Label{L3}
Let $P_{Y|X_1 X_2}$ be a MAC with two input systems.
Assume that the $\sN$ messages are randomly selected as 
the random variables
$X_{2,1},\ldots, X_{2,\sN}$ that are independently generated subject to $P_{X_2}$.
Also, assume that the receiver knows the other random variable $X_1$
and selects it subject to $P_{X_1}$ independently of
$X_{2,1},\ldots, X_{2,\sN}$.
Here, we denote the above selection by ${\cC'}$ and 
the decoding error probability under the maximum likelihood (ML) decoder
by $e_{\cC'}$.
Then, for $s \in [0,1]$, we have
\begin{align}
\mathbb{E}_{\cC'} 
\left[e_{\cC'}\right]
\le  
\sN^s
e^{-s I_{\frac{1}{1+s}}^{\uparrow} (Y X_{1};  X_{2})}
=\sN^s
e^{-s I_{\frac{1}{1+s}}^{\uparrow} (Y ;  X_{2}| X_{1})}.\Label{NMD}
\end{align}
\end{lemma}
\begin{IEEEproof}
Lemma \ref{L3} can be easily shown from Gallager's bound \cite{Gallager68}.
The above situation can be considered as the channel coding with the channel given by the conditional distribution
$P_{YX_1|X_2}$.
Application of Gallager's bound to this channel yields \eqref{NMD}.
\end{IEEEproof}

\subsection{Exponential evaluation}\label{sec: Exponential evaluation}
Consider the case where both encoders use non-adaptive codes and both decoders use the 2nd type decoder. 

{\it Codebook Generation:}
At transmitter $i,$ let $V_{i,1}^n,\ldots, V_{i, \sM_i \cdot \sL_i}^n$ 
be random variables that are independently generated subject to 
$P_{V_i^n}=\prod\limits_{j=1}^{n}P_{V_{i,j}}, $ where 
$V_i^n=(V_{i,1}, \cdots, V_{i,n}),$  
$\sM_i= e^{n R_i}$ and 
$\sL_i= e^{n r_i}$ for $i=1,2.$

{\it Encoding:}
When the legitimate user $i$ intends to send the message $m_i$, he(she) chooses one of 
${V}^n_{i,(m_i-1) \sL_i+1}, \ldots, {V}^n_{i,m_i \sL_i}$ with equal probability, which is denoted to be ${V}^n_{i,(m_i, m_{i,r})}$ and sends it. 

{\it Decoding:}
At the other legitimate receiver, an ML decoder will be used to decode ${V}^n_{i,(m_i, m_{i,r})}$ and thus $m_{i}$
by using the receiver's information $X_{i \oplus 1}^n$. 

To show Theorem \ref{Cor: S-MAC-KS}, we prepare the following Lemma,
which gives exponential evaluation on both the decoding error probability  and the information leaked to the eavesdropper.
\begin{lemma}\Label{L3T}
For a joint distribution $P_{V_1 V_2 X_1 X_2} \in {\cal Q}$,
a rate pair $(R_1,R_2)$, and two positive numbers $r_1,r_2$,
there exists a sequence of $(e^{nR_1}, e^{nR_2}, n)$  non-adaptive codes $\Co_n$ with the 2nd type decoder such that
	\begin{align} 
	  P_{e}^n(\Co_n) &\leq \ 2\left(
e^{ns\left(  R_1+r_1-  I_{\frac{1}{1+s}}^{\uparrow} (Y_2 ;  V_{1}| X_{2})\right)}
+e^{ns\left(  R_2+r_2-  I_{\frac{1}{1+s}}^{\uparrow} (Y_1 ;  V_{2}| X_{1})\right)}\right)
, \Label{eq:ReliabilityD} \\
	  I(M_1, M_2;Z^n|\Co_n)	&\leq	2\left(
	  \sum_{\mathcal{S}\neq \emptyset, \mathcal{S}\subset  \{1,2\}}
e^{n s\left( I^{\downarrow}_{1+s} ( Z ; V_{\mathcal{S}})-\sum\limits_{j \in \mathcal{S}}r_j\right) }\right). \Label{eq:SSecD}
\end{align}
Also, there exists a sequence of $(e^{nR_1}, e^{nR_2}, n)$  non-adaptive codes $\Co_n$ with the 2nd type decoder such that
	\begin{align} 
	  P_{e}^n(\Co_n) &\leq 3 \left(e^{ns\left(  R_1+r_1-  I_{\frac{1}{1+s}}^{\uparrow} (Y_2 ;  V_{1}| X_{2})\right)}
+e^{ns\left(  R_2+r_2-  I_{\frac{1}{1+s}}^{\uparrow} (Y_1 ;  V_{2}| X_{1})\right)}\right)
, \Label{eq:Reliability2D} \\
	  I(M_1;Z^n|\Co_n)	&\leq 3 \left(e^{n s\left( I^{\downarrow}_{1+s} ( Z ; V_1,V_2)-(r_1+r_2+R_2)\right) }
+e^{n s\left( I^{\downarrow}_{1+s} ( Z ; V_1)- r_1\right) }
+e^{n s\left( I^{\downarrow}_{1+s} ( Z ; V_2)-(r_2+R_2)\right) }\right)
, \Label{eq:SSec2D} \\
	  I( M_2;Z^n|\Co_n)	&\leq 3\left(
	  e^{n s\left( I^{\downarrow}_{1+s} ( Z ; V_1,V_2)-(r_1+r_2+R_1)\right) }
+e^{n s\left( I^{\downarrow}_{1+s} ( Z ; V_2)- r_2\right) }
+e^{n s\left( I^{\downarrow}_{1+s} ( Z ; V_1)-(r_1+R_1)\right) }\right)
	  . \Label{eq:SSec3D}
\end{align}
\end{lemma}

\begin{IEEEproof}
According to the reliability analysis in Section \ref{S4-B} and the strong secrecy analysis based on resolvability in Section \ref{S4-C}, we have
	\begin{align} 
\mathbb{E}_{\Co_n}	  \left[P_{e}^n(\Co_n)\right] &\leq 
e^{ns\left(  R_1+r_1-  I_{\frac{1}{1+s}}^{\uparrow} (Y_2 ;  V_{1}| X_{2})\right)}
+e^{ns\left(  R_2+r_2-  I_{\frac{1}{1+s}}^{\uparrow} (Y_1 ;  V_{2}| X_{1})\right)}
, \Label{eq:ReliabilityE} \\
\mathbb{E}_{\Co_n}	  \left[I(M_1, M_2;Z^n|\Co_n)\right]	&\leq	
	  \sum_{\mathcal{S}\neq \emptyset, \mathcal{S} \subset  \{1,2\}}
e^{n s\left( I^{\downarrow}_{1+s} ( Z ; V_{\mathcal{S}})-\sum\limits_{j \in \mathcal{S}}r_j\right) }, \Label{eq:SSecE}\\
\mathbb{E}_{\Co_n}	  \left[I(M_1;Z^n|\Co_n)\right]	&\leq 
e^{n s\left( I^{\downarrow}_{1+s} ( Z ; V_1,V_2)-(r_1+r_2+R_2)\right) }
+e^{n s\left( I^{\downarrow}_{1+s} ( Z ; V_1)- r_1\right) }
+e^{n s\left( I^{\downarrow}_{1+s} ( Z ; V_2)-(r_2+R_2)\right) }
, \Label{eq:SSec2E} \\
\mathbb{E}_{\Co_n}	  \left[I( M_2;Z^n|\Co_n)\right]	&\leq 
	  e^{n s\left( I^{\downarrow}_{1+s} ( Z ; V_1,V_2)-(r_1+r_2+R_1)\right) }
+e^{n s\left( I^{\downarrow}_{1+s} ( Z ; V_2)- r_2\right) }
+e^{n s\left( I^{\downarrow}_{1+s} ( Z ; V_1)-(r_1+R_1)\right) }
	  . \Label{eq:SSec3E}
\end{align}
With Markov inequality, 
the relations \eqref{eq:ReliabilityE} and \eqref{eq:SSecE}
imply the first statement of Lemma \ref{L3T}.
In the same way, 
the relations \eqref{eq:ReliabilityE}, \eqref{eq:SSec2E}, 
and \eqref{eq:SSec3E}
imply the second statement of Lemma \ref{L3T}.
\end{IEEEproof}

\subsection{Reliability Analysis Based on Gallager's Bound}\Label{S4-B}
We show \eqref{eq:ReliabilityE}. Consider the case when User 1 decodes the message $M_2,$ while holding $X_1^n=(X_{i,1}, \cdots, X_{i,n})$ which was chosen according to 
the distribution $P_{X_1^n}=\prod\limits_{j=1}^{n}P_{X_{1,j}}$.
User 2 encodes the message 
into one of $V_{2,1}^n,\ldots, V_{2, \sM_2 \cdot \sL_2}^n$ 
that are independently generated subject to 
$P_{V_2^n}=\prod\limits_{j=1}^{n}P_{V_{2,j}}$.

We apply Lemma \ref{L3} to
the $n$-fold asymptotic case
with $\sN_i=\sM_i\sL_i= e^{n (R_i+r_i)}$.
In this case, the relation \eqref{Ne3} guarantees the additivity property
\allowdisplaybreaks
\begin{align}
I_{\frac{1}{1+s}}^{\uparrow} (Y_1^n ;  V_{2}^n| X_{1}^n)
=
n I_{\frac{1}{1+s}}^{\uparrow} (Y_1 ;  V_{2}| X_{1}), \quad
\Label{Ne4}
 \end{align}
Then, we found that 
the expectation of the decoding error probability of User 1 
is upper bounded by 
$e^{ns\left(  R_2+r_2-  I_{\frac{1}{1+s}}^{\uparrow} (Y_1 ;  V_{2}| X_{1})\right)}$.
In the same way, we can show that the expectation of 
the decoding error probability of User 2 
is upper bounded by 
$e^{ns\left(  R_1+r_1-  I_{\frac{1}{1+s}}^{\uparrow} (Y_2 ;  V_{1}| X_{2})\right)}$.
Hence, we obtain \eqref{eq:ReliabilityE}.

\subsection{Strong secrecy analysis based on resolvability}\Label{S4-C}
For the strong secrecy analysis, we consider both joint secrecy and individual secrecy.

First, we consider the joint secrecy, i.e., to show \eqref{eq:SSecE}.
We have
\begin{align}
\mathbb{E}_{\Co_n} \left[I(M_1,M_2;Z^n|\Co_n)\right]=&\mathbb{E}_{\Co_n}\left[\sum\limits_{m_1,m_2}P_{M_1 M_2}
(m_1,m_2)
D(P_{Z^n|M_1=m_1,M_2=m_2} \| P_{Z^n|\Co_n})\right]\nonumber\\
=&
\mathbb{E}_{\Co_n}\left[\sum\limits_{m_1,m_2}
P_{M_1 M_2}
(m_1,m_2)
D(P_{Z^n|M_1=m_1,M_2=m_2} \| P_{Z^n})
-
D( P_{Z^n|\Co_n}\|  P_{Z^n} )\right]
\nonumber\\
\le &
\mathbb{E}_{\Co_n}\left[\sum\limits_{m_1,m_2}
P_{M_1 M_2}
(m_1,m_2)
D(P_{Z^n|M_1=m_1,M_2=m_2} \| P_{Z^n})\right] \nonumber\\
=&\sum\limits_{m_1,m_2}
P_{M_1 M_2} \ 
(m_1,m_2)
\mathbb{E}_{\Co_n}\left[D(P_{Z^n|M_1=m_1,M_2=m_2} \| P_{Z^n})\right], 
\Label{MVR}
\end{align} 
where $\mathcal{M}_1, \mathcal{M}_2$
are fixed sets of messages, and correspondingly
\begin{align*}
P_{Z^n|M_1=m_1,M_2=m_2}(z^n)&:=
\frac{1}{\sL_1 \sL_2} 
\sum_{j_1,j_2} P_{Z^n| V_1^n=V^n_{1,j_1}, V^n_2=V^n_{2,j_2}  }(z^n),
\end{align*}
with $(m_i-1) \sL_i+1\leq j_i\leq m_i \sL_i,$ for $1\leq i\leq k.$

We apply Lemma \ref{L2} to the $n$-fold asymptotic case
with inputs $(V_1^n, V_2^n)$ and $\sL_i= e^{n r_i}$.
In this case, the relation \eqref{Ne1} guarantees the additivity property
\begin{align}
 I^{\downarrow}_{1+s}(Z^n; V^n_{\mathcal{S}})=n I^{\downarrow}_{1+s}(Z; V_{\mathcal{S}}), \quad \mbox{for } \mathcal{S}\neq \emptyset, \mathcal{S}\subset  \{1,2\}.
\Label{Ne2}
 \end{align}
Lemma \ref{L2} and \eqref{Ne2} together
imply that
\begin{align}
\mathbb{E}_{\cC_n} \left[D(  P_{Z^n|M_1=m_1,M_2=m_2}\| P_{Z^n})\right] \le 
\sum_{\mathcal{S}\neq \emptyset, \mathcal{S}\subset  \{1,2\}}
e^{n s\left( I^{\downarrow}_{1+s} ( Z ; V_{\mathcal{S}})-\sum\limits_{j \in \mathcal{S}}r_j\right) }.
\Label{NBT}
\end{align}
The combination of \eqref{MVR} and \eqref{NBT}
implies \eqref{eq:SSecE}.

Next, we consider the individual secrecy, i.e., to show \eqref{eq:SSec2E}.
We apply Lemma \ref{L2} to
the $n$-fold asymptotic case
with $\sN_i=\sM_i\sL_i= e^{n (R_i+r_i)}$.
\begin{align}
\mathbb{E}_{\Co_n} \left[ I(M_i ;Z^n|\Co_n)\right]=&
\mathbb{E}_{\Co_n}\left[\sum\limits_{m_i}P_{M_i}
(m_i)
D(P_{Z^n|M_i=m_i} \| P_{Z^n|\Co_n})\right]\nonumber \\
\le &
\sum\limits_{m_i}
P_{M_i} (m_i)
\mathbb{E}_{\Co_n}\left[D(P_{Z^n|M_i=m_i} \| P_{Z^n})\right], \Label{MVR2}
\end{align} 
Then, we have
\begin{align*}
P_{Z^n|M_1=m_1}(z^n)&:=
\frac{1}{\sL_1 \sN_2} 
\sum_{j_1=(m_i-1) \sL_i+1}^{m_i \sL_i}
\sum_{j_2=1}^{\sN_2} P_{Z^n| V_1^n=V^n_{1,j_1}, V^n_2=V^n_{2,j_2}  }(z^n),
\end{align*}
where $\sL_1= e^{n r_1}$ and $\sN_2= e^{n (R_2+r_2)}.$ 

Lemma \ref{L2} and \eqref{Ne2} together
implies that
\begin{align}
\mathbb{E}_{\cC_n} \left[D(  P_{Z^n|M_1=m_1}\| P_{Z^n})\right] \le 
e^{n s\left( I^{\downarrow}_{1+s} ( Z ; V_1,V_2)-(r_1+r_2+R_2)\right) }
+e^{n s\left( I^{\downarrow}_{1+s} ( Z ; V_1)- r_1\right) }
+e^{n s\left( I^{\downarrow}_{1+s} ( Z ; V_2)-(r_2+R_2)\right) }\Label{MVR3}.
\end{align}
The relations \eqref{MVR2} and \eqref{MVR3} imply \eqref{eq:SSec2E}.
In the same way, we can show \eqref{eq:SSec3E}.

\subsection{Deriving the achievable region under strong secrecy}\Label{S4-D}
In order to show the relations \eqref{NZT},
it is sufficient to  
show the relations
\begin{align}
{\cal R}_{J,N}(P_{V_1 V_2 X_1 X_2}) \subset {\cal C}_{J,S,N,2},\quad &
{\cal R}_{I,N}(P_{V_1 V_2 X_1 X_2}) \subset {\cal C}_{I,S,N,2}\Label{NZTV} 
\end{align}
for $P_{V_1 V_2 X_1 X_2}\in {\cal Q}$,
because the application of time sharing to the obtained region in 
\eqref{NZTV} yields the relations \eqref{NZT}.
Therefore, in the following, we show the relations 
\eqref{NZTV}.

Using Lemma \ref{L3T}, we  prove Theorem \ref{Cor: S-MAC-KS}.
When 
\begin{align}
R_2+r_2 &< I(Y_1;V_2|X_1), \Label{G5}\\
R_1+r_1 &< I(Y_2;V_1|X_2),\Label{G4}
\end{align}
the decoding error probability $P_{e}^n(\Co_n) $ goes to zero as $n\to \infty$.

First, we address the strong joint secrecy.
When 
\begin{align}
\sum_{i \in \mathcal{S}} r_i > I ( Z ; V_{\mathcal{S}}), \quad \forall \mathcal{S}\neq \emptyset, \mathcal{S}\subseteq 
\{1,2\},
 \Label{E1}
\end{align}
the mutual information $ I(M_1, M_2;Z^n|\Co_n)$ goes to zero as $n\to \infty$.
For strong joint secrecy, 
we have the following constraints on rates $r_i$ for the two-way wiretap channel:
\begin{align}
	r_1 &> I(Z;V_1); \Label{G1Y}\\
	r_2 &> I(Z;V_2);\Label{G2Y}\\
	r_1+r_2 &> I(Z;V_1, V_2).\Label{G3Y}
\end{align}
For the joint secrecy,
applying Fourier-Motzkin elimination to remove $r_1,r_2$ 
in \eqref{G5} -- \eqref{G3Y},
we obtain the rate region of $(R_1,R_2)$ defined by
\begin{align}
\begin{split}
	R_1 &\leq I(Y_2;V_1|X_2)-I(Z;V_1)\\
	R_2 &\leq I(Y_1;V_2|X_1)-I(Z;V_2)\\
	R_1+R_2 & \leq  I(Y_2;V_1|X_2) + I(Y_1;V_2|X_1) - I(V_1, V_2;Z),
\end{split}\Label{HH1}
\end{align}
which implies the first relation in \eqref{NZT}.

Next, we address the strong individual secrecy.
When 
\begin{align*}
r_1 &> I( Z ; V_1), \\
r_2+R_2 &> I( Z ; V_2),\\
r_1+r_2+R_2 &> I( Z ; V_1,V_2), 
\end{align*}
the mutual information $ I(M_1;Z^n|\Co_n)$ goes to zero as $n\to \infty$.
Since the same discussion holds for 
the mutual information $ I(M_2;Z^n|\Co_n)$
for strong individual secrecy, 
we have the following constraints on rates $r_i$ and $R_i$ 
for the two-way wiretap channel:
\begin{align}
	r_1 &> I(Z;V_1);\Label{G1}\\
	r_2 &> I(Z;V_2);\Label{G2}\\
	r_1+r_2+\min(R_1,R_2) &> I(Z;V_1, V_2).\Label{G3}
\end{align}

For the individual secrecy,
applying Fourier-Motzkin elimination to remove $r_1,r_2$ 
in \eqref{G5}, \eqref{G4}, and \eqref{G1} -- \eqref{G3},
we obtain the rate region of $(R_1,R_2)$ defined by
\begin{align}
\begin{split}
R_1 &\leq I(Y_2;V_1|X_2)-I(Z;V_1)\\
R_2 &\leq I(Y_1;V_2|X_1)-I(Z;V_2)\\
\max\{R_1, R_2\} & \leq  I(Y_2;V_1|X_2) + I(Y_1;V_2|X_1) - I(V_1, V_2;Z),
\end{split}\Label{HH2}
\end{align}
which implies the second relation in \eqref{NZT}. The detail of Fourier-Motzkin elimination is explained in Appendix \ref{AP1}.

\section{Proof of Theorem \ref{Cor: C-MAC-KS}}\Label{sec: Proof2}
To prove Theorem \ref{Cor: C-MAC-KS}, we need to find a sequence of non-adaptive codes ${\mathcal{C}_n}$ with peak cost constraint $c_1, c_2$ such that
\begin{itemize}
	\item the codes have rates pair converge to $(R_1, R_2)$ within the region ${\cal R}_{J,N}^{c_1, c_2}$ and ${\cal R}_{I,N}^{c_1, c_2}$ for joint secrecy and individual secrecy, respectively;
	\item the decoding error is bounded by $\epsilon_n$ with $\lim\limits_{n\to \infty} \epsilon_n =0;$
	\item the information leakage is bounded by $\tau_n$ with $\lim\limits_{n\to \infty} \tau_n =0.$
\end{itemize}
To this end, we use constant composition codes in our coding scheme to satisfy the cost constraint and evaluate the decoding error (upper bounded by $\epsilon_n$) and respective information leakage (upper bounded by $\tau_n$) for joint secrecy and individual security (similar to the proof of Theorem \ref{Cor: S-MAC-KS}). As a preparation, we first introduce the basics of type method in Sec. \ref{SV-1}, which is used to establish the constant composition version of the resolvability lemma and generalization of Gallager's bound in Sec. \ref{sec: Constant composition codes}. The coding scheme using constant composition codes and its analysis are provided in Sec. \ref{S6-1}. Sec. \ref{sec: Deriving the achievable region} derives the desired region in Theorem \ref{Cor: C-MAC-KS}.   

\subsection{Basics of type method}\Label{SV-1}
First, we give notations for the method of types for a single finite system ${\cal X}$ with $d_X=|{\cal X}|$.
Given an element $x^n \in {\cal X}^n$ and an element $x\in {\cal X}$, 
we define the subset ${\cal N}(x^n,x) := \{ i| x_i=x\} $, 
the integer $n_x :=| {\cal N}(x^n,x) |$,
and the empirical distribution 
$T_Y(x^n):= (\frac{n_1}{n},\ldots, \frac{n_{d_X}}{n})$, which is called a type.
The set of types is denoted by $T_n({\cal X})$.
For ${P} \in T_n({\cal X})$, a subset of ${\cal X}^n$ is defined by:
\begin{align*}
&T_{P}^n({\cal X})
:= \{x^n \in {\cal X}^n| 
T_Y(x^n)=P\}.
\end{align*}
We call $T_{P}^n({\cal X})$ the fixed-type subset of $P$.
We simplify $T_{P}^n({\cal X})$ to $T_P^n$ when we do not need to identify  ${\cal X}.$  
In this paper, the value $\nu_n( d_X):=\max\limits_{P\in T_n({\cal X})} \frac{e^{nH(P)}}{|T_P^n|} \le (1+n)^{d_X}$
plays a key role.
The uniform distribution $P_{\Unif,T_P^n}$ on the subset $T_P^n$ satisfies
\begin{align}
 P_{\Unif,T_P^n}  \le \nu_n( d_X)   P^{n}
\le (1+n)^{d_X}  P^{n}
\Label{eq3-X}.
\end{align}

Next, we discuss the method of types for two finite systems ${\cal X}$ 
and ${\cal V}$.
For $P_{XV} \in T_n({\cal X}\times {\cal V})$ and $v^n \in {\cal V}^n$, 
we define  a subset of $ {\cal X}^n$ as
\begin{align*}
&T_{P_{X|V}}^n({\cal X},v^n)
:= \{x^n \in {\cal X}^n| 
(x^n,v^n) \in T_{P_{XV}}^n\}.
\end{align*}
We simplify $T_{P_{X|V}}^n({\cal X},v^n)$ to $T_{P_{X|V}}^n(v^n)$.

Now, we introduce notations over the finite sets
${\cal X}_1$, ${\cal X}_2$, ${\cal V}_1$ and ${\cal V}_2.$
We fix a positive integer $n$.
Let $P_{Z|X_1 X_2}$ be a MAC with two inputs system.
Let $P_{V_1 X_1}$ and $P_{V_2 X_2}$ be joint types with length $n$.
Define 
$P_{Z|V_1 V_2}(z|v_1,v_2):= \sum\limits_{x_1,x_2}P_{Z|X_1 X_2}(z|x_1,x_2)
P_{X_1|V_1}(x_1|v_1) P_{X_2|V_2}(x_2|v_2)$
and
$P_{V_1 V_2}:=P_{V_1}\times P_{V_2}$.
Then, we have the joint distribution $P_{Z V_1 V_2}$.
Based on this joint distribution, we define 
the information quantity $\breve{I}^{\uparrow}_{\frac{1}{1-s}} ( Z ; V_{\mathcal{S}})$
for $\mathcal{S}\neq \emptyset, \mathcal{S}\subset \{1,2\}$.

Define the uniform distribution $\breve{P}_{X_i^n V_i^n}$ on 
the subset $T_{P_{V_i X_i}}^n $ of 
$({\cal V}_i \times {\cal X}_i)^n$ for $i=1,2$.
Then, we define its marginal distributions 
$\breve{P}_{V_i^n}$,
$\breve{P}_{X_i^n}$
and its conditional distribution $\breve{P}_{X_i^n |V_i^n}$.
Notice that 
$\breve{P}_{V_i^n}$ and $\breve{P}_{X_i^n}$ are the  uniform distributions on 
the subsets 
$T_{P_{V_i}}^n $ and
$T_{P_{X_i}}^n $, respectively.
Also, 
the conditional distribution $\breve{P}_{X_i^n |V_i^n=v_i^n}$
is the uniform distribution on $T^n_{P_{X_i|V_i}}(v^n)$.

\subsection{Constant composition codes}\label{sec: Constant composition codes}
We define two-transmitter MAC 
as the conditional distribution $P_{Z^n|V_1^n V_2^n}:=
P_{Z^n|X_1^n X_2^n}\cdot
\breve{P}_{X_1^n|V_1^n} \cdot \breve{P}_{X_2^n|V_2^n}$.
Then, we have a constant composition version of 
the following resolvability lemma.
\begin{lemma}\Label{L2C}
Let $V_{i,1}^n,\ldots, V_{i,\sL_i}^n$ 
be the random variables that are independently generated subject to $\breve{P}_{V_i^n}$ for $i=1,2$.
Here, we denote this code selection by ${\cC}$ and define
the distribution $P_{Z^n| \cC}$ for the variable $Z^n$ dependently of 
the code selection ${\cC}$ as
$P_{Z^n| \cC}(z^n):=
\frac{1}{\sL_1 \sL_2} 
\sum\limits_{j_1,j_2} P_{Z^n| V_1^n=V_{1,j_1}^n, V_2^n=V_{2,j_2}^n  }(z^n)$.

Then, 
for $s \in [0,1]$, we have $ 
D(  P_{Z^n| \cC}\| P_{Z^n}) 
\le  
D_{1+s}(  P_{Z^n| \cC}\| P_{Z^n})$ and
\begin{align}
s \mathbb{E}_{\cC} 
\left[ D_{1+s}(  P_{Z^n| \cC}\| P_{Z^n})\right] 
\le 
\sum_{\mathcal{S}\neq \emptyset, \mathcal{S}\subset \{1,2\}}
\frac{\nu_n(|{\cal X}_1|)\nu_n(|{\cal X}_2|)
}{ \prod_{i \in \mathcal{S}} \sL_i^s} 
e^{s \breve{I}_{\frac{1}{1-s}}^{\uparrow} ( Z ; V_{\mathcal{S}}) }\Label{NND}
\end{align}
for $s \in [0,1]$.
\end{lemma}
\begin{IEEEproof}
	See Appendix \ref{A-F} for a detailed proof.
\end{IEEEproof}

To state a generalization of Gallager's bound \cite{Gallager68} 
for constant composition codes,
we assume that $|{\cal Y}|$, $|{\cal X}_1|$, $|{\cal X}_2|$, and $|{\cal V}_2|$ are finite.
We fix a positive integer $n$.
Let $P_{Y|X_1 X_2}$ be a MAC with two inputs system.
Let $P_{X_1}$ and $P_{V_2 X_2}$ be (joint) types with length $n$.
Define 
$P_{Y|X_1 V_2}(y|x_1,v_2):= \sum\limits_{x_2}P_{Y|X_1 X_2}(y|x_1,x_2)
P_{X_2|V_2}(x_2|v_2)$
and
$P_{X_1 V_2}:=P_{X_1}\times P_{V_2}$.
Then, we have the joint distribution $P_{Y X_1 V_2}$.
Based on this joint distribution, we define 
the information quantity 
$\breve{I}_{\frac{1}{1+s}}^{\uparrow} (Y ;  V_{2}| X_{1})$.

Define the uniform distribution $\breve{P}_{X_2^n V_2^n}$ on 
the subset $T_{P_{V_2 X_2}}^n $ of 
$({\cal V}_2 \times {\cal X}_2)^n.$ 
Then, we define its marginal distribution $\breve{P}_{V_2^n}$
and its conditional distribution $\breve{P}_{X_2^n |V_2^n}$.
We define the uniform distribution $\breve{P}_{X_1^n}$ on 
the subset $T_{P_{X_1}}^n $ of ${\cal X}_1^n$.
We define two-transmitter MAC
$P_{Y^n|X_1^n V_2^n}(y^n|x_1^n,v_2^n):=
\sum\limits_{x_2^n}
P_{Y^n|X_1^n X_2^n}(y^n|x_{1}^n,x_{2}^n)
\breve{P}_{X_2^n|V_2^n}(x_{2}^n|v_{2}^n)$.

Then, we have the following as a generalization of Gallager's bound \cite{Gallager68}. 
\begin{lemma}\Label{L3C}
Assume that the $\sN$ messages are randomly selected as 
the random variables
$V_{2,1}^n,\ldots, V_{2,\sN}^n$ that are independently generated subject to 
$\breve{P}_{V_2^n}$.
Also, assume that the receiver knows the other random variable $X_1^n$
and selects it subject to $\breve{P}_{X_1^n}$ independently of
$V_{2,1}^n,\ldots, V_{2,\sN}^n$.
Here, we denote the above selection by ${\cC'}$ and 
the decoding error probability under the maximum likelihood (ML) decoder
by $e_{\cC'}$.

Then, for $s \in [0,1]$, we have
\begin{align}
\mathbb{E}_{\cC'} 
\left[ e_{\cC'}\right]
\le 
	|T_n({\cal Y}\times {\cal X}_1)|^{1+s} \nu_n(|{\cal Y}| |{\cal X}_1| )^{s} 
	\nu_n(|{\cal X}_1|)\nu_n(|{\cal X}_2|) 
\sN^s
e^{-s \breve{I}_{\frac{1}{1+s}}^{\uparrow} (Y ;  V_{2}| X_{1})}.
\Label{ZMH}
\end{align}
\end{lemma}
\begin{IEEEproof}
	See Appendix \ref{A-G} for a detailed proof.
\end{IEEEproof}

\subsection{Exponential evaluation}\Label{S6-1}
To show Theorem \ref{Cor: C-MAC-KS}, we prepare the following Lemma,
which gives exponential evaluation.
\begin{lemma}\Label{L3TC}
Assume that $|{\cal Y}_1|$, $|{\cal Y}_2|$, $|{\cal X}_1|$, $|{\cal X}_2|$, $|{\cal V}_1|$, and $|{\cal V}_2|$ 
are finite.
We fix a positive integer $n$.
Let $P_{Z Y_1 Y_2|X_1 X_2}$ be a MAC with two inputs system.
Let $P_{V_1 X_1}$ and $P_{V_2 X_2}$ be joint types with length $n$.
Define 
$$ P_{Z Y_1 Y_2 V_1 V_2 X_1 X_2}(z,v_1,v_2,x_1,x_2):= 
P_{Z Y_1 Y_2|X_1 X_2}(z,y_1,y_2|x_1,x_2)
P_{X_1 V_1}(x_1,v_1) P_{X_2 V_2}(x_2,v_2).$$
Then, based on this joint distribution, we define 
the information quantities
$\breve{I}_{\frac{1}{1+s}}^{\uparrow} (Y_2 ;  V_{1}| X_{2})$,
$\breve{I}_{\frac{1}{1+s}}^{\uparrow} (Y_1 ;  V_{2}| X_{1})$, and
 $\breve{I}_{\frac{1}{1-s}}^{\uparrow} ( Z ; V_{\mathcal{S}})$
for $\mathcal{S}\neq \emptyset, \mathcal{S}\subset \{1,2\}$.
We choose $\nu_n$ and $\nu_n'$ as
\begin{align}
\nu_n:=&\nu_n(|{\cal X}_1||{\cal X}_2| \max( |{\cal V}_{1}|, |{\cal V}_{2}|)),\\
\nu_n':= &\nu_n(|{\cal X}_1|)\nu_n(|{\cal X}_2|) 
\max(	|T_n({\cal Y}\times {\cal X}_1)|^{1+s} \nu_n(|{\cal Y}| |{\cal X}_1| )^{s}, 
	|T_n({\cal Y}\times {\cal X}_2)|^{1+s} \nu_n(|{\cal Y}| |{\cal X}_2| )^{s} ).
	\end{align}

For 
a rate pair $(R_1,R_2)$, and two positive numbers $r_1,r_2$,
there exists a sequence of $(e^{nR_1}, e^{nR_2}, n)$ 
non-adaptive codes $\Co_n$ with the 2nd type decoder
in $T_{P_{X_1}}^n \times T_{P_{X_2}}^n $ 
such that
	\begin{align} 
	  P_{e}^n(\Co_n) &\leq \ 2 \nu_n'\left(
e^{ns\left(  R_1+r_1-  \breve{I}_{\frac{1}{1+s}}^{\uparrow} (Y_2 ;  V_{1}| X_{2})\right)}
+e^{ns\left(  R_2+r_2-  \breve{I}_{\frac{1}{1+s}}^{\uparrow} (Y_1 ;  V_{2}| X_{1})\right)}\right)
, \Label{eq:ReliabilityDC} \\
	  I(M_1, M_2;Z^n|\Co_n)	&\leq	2 \nu_n \left(
	  \sum_{\mathcal{S}\neq \emptyset, \mathcal{S}\subset  \{1,2\}}
e^{n s\left( \breve{I}_{\frac{1}{1-s}}^{\uparrow} ( Z ; V_{\mathcal{S}})-\sum\limits_{j \in \mathcal{S}}r_j\right) }\right). \Label{eq:SSecDC}
\end{align}
Also, there exists a sequence of $(e^{nR_1}, e^{nR_2}, n)$ non-adaptive codes $\Co_n$ 
in $T_{P_{X_1}}^n \times T_{P_{X_2}}^n $ 
such that
	\begin{align} 
	  P_{e}^n(\Co_n) &\leq 3 \nu_n' \left(e^{ns\left(  R_1+r_1-  \breve{I}_{\frac{1}{1+s}}^{\uparrow} (Y_2 ;  V_{1}| X_{2})\right)}
+e^{ns\left(  R_2+r_2-  \breve{I}_{\frac{1}{1+s}}^{\uparrow} (Y_1 ;  V_{2}| X_{1})\right)}\right)
, \Label{eq:Reliability2DC} \\
	  I(M_1;Z^n|\Co_n)	&\leq 3 \nu_n \left(e^{n s\left( \breve{I}_{\frac{1}{1-s}}^{\uparrow}( Z ; V_1,V_2)-(r_1+r_2+R_2)\right) }
+e^{n s\left( \breve{I}_{\frac{1}{1-s}}^{\uparrow} ( Z ; V_1)- r_1\right) }
+e^{n s\left( \breve{I}_{\frac{1}{1-s}}^{\uparrow} ( Z ; V_2)-(r_2+R_2)\right) }\right)
, \Label{eq:SSec2DC} \\
	  I( M_2;Z^n|\Co_n)	&\leq 3 \nu_n \left(
	  e^{n s\left( \breve{I}_{\frac{1}{1-s}}^{\uparrow} ( Z ; V_1,V_2)-(r_1+r_2+R_1)\right) }
+e^{n s\left( \breve{I}_{\frac{1}{1-s}}^{\uparrow} ( Z ; V_2)- r_2\right) }
+e^{n s\left( \breve{I}_{\frac{1}{1-s}}^{\uparrow} ( Z ; V_1)-(r_1+R_1)\right) }\right)
	  . \Label{eq:SSec3DC}
\end{align}
\end{lemma}

\begin{IEEEproof}
The transmitters encode their messages independently. 
Before transmission, they generate
the random variables 
$V_{i,1}^n,\ldots, V_{i, \sM_i \cdot \sL_i}^n$ 
independently subject to 
$\breve{P}_{V_i^n},$
where 
$\sM_i= e^{n R_i}$ and 
$\sL_i= e^{n r_i}$ for $i=1,2$.

{\it Encoding:}
When the legitimate user $i$ intends to send the message $m_i$, 
he(she) chooses one of  ${V}^n_{i,(m_i-1) \sL_i+1}, \ldots, {V}^n_{i,m_i \sL_i}$.
When the selected one element is $v_i^n$,
user $i$ generates $x_i^n$ according to $\breve{P}_{X_i^n |V_i^n=v_i^n}$, 
and sends $x_i^n$ to the channel. 

{\it Decoding:}
At the other legitimate receiver, a ML decoder will be used to decode $M_{i}$
by using the receiver's information $X_{i \oplus 1}^n$. 

Then, in the same way as the proof of Lemma \ref{L3T},
replacing the roles of Lemmas \ref{L2} and \ref{L3}
by the roles of Lemmas \ref{L2C} and \ref{L3C},
we have
	\begin{align} 
\mathbb{E}_{\Co_n}	  \left[ P_{e}^n(\Co_n) \right]&\leq 
\nu_n'\left(e^{ns\left(  R_1+r_1-  \breve{I}_{\frac{1}{1+s}}^{\uparrow} (Y_2 ;  V_{1}| X_{2})\right)}
+e^{ns\left(  R_2+r_2-  \breve{I}_{\frac{1}{1+s}}^{\uparrow} (Y_1 ;  V_{2}| X_{1})\right)}\right)
, \Label{eq:ReliabilityEC} \\
\mathbb{E}_{\Co_n}	 \left[ I(M_1, M_2;Z^n|\Co_n)\right]	&\leq	
\nu_n\left(	  \sum_{\mathcal{S}\neq \emptyset, \mathcal{S}\subset  \{1,2\}}
e^{n s\left( \breve{I}_{\frac{1}{1-s}}^{\uparrow} ( Z ; V_{\mathcal{S}})-\sum\limits_{j \in \mathcal{S}}r_j\right) }\right), \Label{eq:SSecEC}\\
\mathbb{E}_{\Co_n}	 \left[ I(M_1;Z^n|\Co_n)\right]	&\leq \nu_n\left(
e^{n s\left( \breve{I}_{\frac{1}{1-s}}^{\uparrow} ( Z ; V_1,V_2)-(r_1+r_2+R_2)\right) }
+e^{n s\left( \breve{I}_{\frac{1}{1-s}}^{\uparrow} ( Z ; V_1)- r_1\right) }
+e^{n s\left( \breve{I}_{\frac{1}{1-s}}^{\uparrow} ( Z ; V_2)-(r_2+R_2)\right) }\right)
, \Label{eq:SSec2EC} \\
\mathbb{E}_{\Co_n}	 \left[  I( M_2;Z^n|\Co_n)\right]	&\leq \nu_n\left(
	  e^{n s\left( \breve{I}_{\frac{1}{1-s}}^{\uparrow} ( Z ; V_1,V_2)-(r_1+r_2+R_1)\right) }
+e^{n s\left( \breve{I}_{\frac{1}{1-s}}^{\uparrow} ( Z ; V_2)- r_2\right) }
+e^{n s\left( \breve{I}_{\frac{1}{1-s}}^{\uparrow} ( Z ; V_1)-(r_1+R_1)\right) }\right)
	  . \Label{eq:SSec3EC}
\end{align}
In the same way as Lemma \ref{L3T},
the above relations yield the statement of Lemma \ref{L3TC}.
\end{IEEEproof}

\subsection{Deriving the achievable region under strong secrecy}\label{sec: Deriving the achievable region}
Assume that $|{\cal Y}_1|$, $|{\cal Y}_2|$, $|{\cal X}_1|$, $|{\cal X}_2|$, $|{\cal V}_1|$, and $|{\cal V}_2|$ 
are finite.
Given $P_{V_1V_2X_1X_2} \in {\cal Q}^{c_1,c_2}$,
we have $T^n_{P_{X_1}}\subset {\cal X}_1^n(c_1)$ and 
$T^n_{P_{X_2}}\subset {\cal X}_2^n(c_2)$. Hence,
in the same way as the discussion in Section \ref{S4-D},
Lemma \ref{L3TC} yields 
\begin{align}
{\cal R}_{J,N}(P_{V_1V_2X_1X_2}) \subset {\cal C}_{J,S,N,2}^{P,c_1,c_2},\quad
{\cal R}_{I,N}(P_{V_1V_2X_1X_2}) \subset {\cal C}_{I,S,N,2}^{P,c_1,c_2}.
\label{NN67}
\end{align}
Applying the time sharing in the same way as Subsection \ref{S4-D},
we obtain the statement of Theorem \ref{Cor: C-MAC-KS}.

Next, we consider the case when $|{\cal Y}_1|$, $|{\cal Y}_2|$, $|{\cal X}_1|$, $|{\cal X}_2|$, $|{\cal V}_1|$, and $|{\cal V}_2|$ 
are not finite.
In the following, we consider the finite-discretization
of ${\cal Y}_1$ and ${\cal Y}_2$.
User $i$ divides the set ${\cal Y}_i$ into finite disjoint subsets 
${\cal R}_i:=\{{\cal T}_{i,1}, \ldots, {\cal T}_{i,l_i}\}$.
Let $\tilde{Y}_i$ be the information for which subset contains $Y_i$. 
Let ${\cal R}_{J,N}(P_{V_1V_2X_1X_2})[{\cal R}_1,{\cal R}_2]$
and
${\cal R}_{I,N}(P_{V_1V_2X_1X_2})[{\cal R}_1,{\cal R}_2]$
be the regions defined in \eqref{BPW1} and \eqref{BPW2} with replacing $Y_i$ by $\tilde{Y}_i$, 
respectively.
The above regions shows the regions when 
User $i$ uses the information $\tilde{Y}_i$ instead of $Y_i$.
When $P_{V_1V_2X_1X_2}' \in {\cal Q}^{c_1,c_2}$ has a finite support,
the relation \eqref{NN67} shows that 
\begin{equation}
{\cal R}_{J,N}(P_{V_1V_2X_1X_2}')[{\cal R}_1,{\cal R}_2]\subset {\cal C}_{J,S,N,2}^{P,c_1,c_2} 
\quad \mbox{and} \quad
 {\cal R}_{I, N}(P_{V_1V_2X_1X_2}')[{\cal R}_1,{\cal R}_2]\subset {\cal C}_{I,S,N,2}^{P,c_1,c_2}.\label{NN68}
\end{equation}

Given $P_{V_1V_2X_1X_2} \in {\cal Q}^{c_1,c_2}$,
we can choose 
finite-discretizations ${\cal R}_1,{\cal R}_2$ 
such that
$\allowbreak {\cal R}_{J,N}(P_{V_1V_2X_1X_2})[{\cal R}_1,{\cal R}_2]$ 
and ${\cal R}_{I, N}(P_{V_1V_2X_1X_2})[{\cal R}_1,{\cal R}_2]$ 
are sufficiently close to  
${\cal R}_{J,N}(P_{V_1V_2X_1X_2})$ and 
$\allowbreak {\cal R}_{I, N}(P_{V_1V_2X_1X_2}),$ respectively.
Also, 
we can choose a distribution $P_{V_1V_2X_1X_2}' \in {\cal Q}^{c_1,c_2}$
with finite support such that
$\allowbreak {\cal R}_{J,N}(P_{V_1V_2X_1X_2}')[{\cal R}_1,{\cal R}_2]$ 
and ${\cal R}_{I, N}(P_{V_1V_2X_1X_2}')[{\cal R}_1,{\cal R}_2]$ 
are sufficiently close to  
$\allowbreak {\cal R}_{J,N}(P_{V_1V_2X_1X_2})[{\cal R}_1,{\cal R}_2]$ 
and ${\cal R}_{I, N}(P_{V_1V_2X_1X_2})[{\cal R}_1,{\cal R}_2]$, respectively.
Therefore, 
given $P_{V_1V_2X_1X_2} \in {\cal Q}^{c_1,c_2}$,
we can choose a pair of 
a distribution $P_{V_1V_2X_1X_2}' \in {\cal Q}^{c_1,c_2}$
with finite support 
and finite-discretizations ${\cal R}_1,{\cal R}_2$ 
such that
$\allowbreak {\cal R}_{J,N}(P_{V_1V_2X_1X_2}')[{\cal R}_1,{\cal R}_2]$ 
and ${\cal R}_{I, N}(P_{V_1V_2X_1X_2}')[{\cal R}_1,{\cal R}_2]$ 
are sufficiently close to  
${\cal R}_{J,N}(P_{V_1V_2X_1X_2})$ and 
$\allowbreak {\cal R}_{I, N}(P_{V_1V_2X_1X_2}),$ respectively.

We denote the inner of a set $S$ by $\inte(S)$.
Due to the above discussion, 
for any element $(R_1,R_2)$ of $\inte({\cal R}_{J,N}(P_{V_1V_2X_1X_2}))$,
there exists
a pair of 
a distribution $P_{V_1V_2X_1X_2}' \in {\cal Q}^{c_1,c_2}$
with finite support 
and finite-discretizations ${\cal R}_1,{\cal R}_2$ 
such that $(R_1,R_2)$ is included in 
$\allowbreak {\cal R}_{J,N}(P_{V_1V_2X_1X_2}')[{\cal R}_1,{\cal R}_2]$.
The combinaiton of this discussion and \eqref{NN68} implies that
$\inte({\cal R}_{J,N}(P_{V_1V_2X_1X_2}))\subset {\cal C}_{J,S,N,2}^{P,c_1,c_2}$. 
In the same way, we can show the relation
$\inte({\cal R}_{I,N}(P_{V_1V_2X_1X_2}))\subset {\cal C}_{I,S,N,2}^{P,c_1,c_2}$.
Since Lemma \ref{LL1} guarantees that 
${\cal C}_{J,S,N,2}^{P,c_1,c_2}$  and ${\cal C}_{I,S,N,2}^{P,c_1,c_2}$ 
are closed sets, we have
${\cal R}_{J,N}(P_{V_1V_2X_1X_2})\subset {\cal C}_{J,S,N,2}^{P,c_1,c_2}$ 
and
${\cal R}_{I,N}(P_{V_1V_2X_1X_2})\subset {\cal C}_{I,S,N,2}^{P,c_1,c_2}$, respectively.
Therefore, we obtain the statement of Theorem \ref{Cor: C-MAC-KS} in the general case.

\section{Additive two-way wiretap channel over a finite field}\label{Sec: Additive TW-WC}
In additive two-way wiretap channel, 
$X_1,X_2,Y_1,Y_2,Z$ are variables in the finite field $\FF_q$.
Then, we have
\begin{align}
Y_1&= a_1 X_1+b_1 X_2+N_1, \Label{AD1}\\
Y_2&= a_2 X_1+ b_2 X_2+N_2, \Label{AD2}\\
Z&=  a_3 X_1+b_3 X_2+N_3, \Label{AD3}
\end{align}
where $a_1,b_1,a_2,b_2,a_3,b_3$ are non-zero elements of $\FF_q$.
$N_1$, $N_2$, $N_3$ are the independent 
random variables in $\FF_q,$ which are independent of $X_1, X_2$.

When $X_1$ and $X_2$ are independent uniform random variables, we have
\begin{align}
I(Y_2;X_1|X_2)&=\log q -H(N_2) \\
I(Y_1;X_2|X_1)&=\log q -H(N_1) \\
I(Z;X_1)&=I(Z;X_2)=0 \\
I(X_1, X_2;Z)&=\log q -H(N_3),
\end{align}
where $H(X)$ expresses the Shannon entropy of a random variable $X$
defined as 
$-\sum\limits_{x}P_X(x) \log P_X(x)$.

Therefore, the joint secrecy achievable rate region \eqref{HH1} is 
evaluated to
 \begin{align}
{\cal R}_{J, N}\supseteq {\cal A}_J:=
\left\{(R_1,R_2)\left|
\begin{array}{l}
	R_1 \leq \log q- H(N_2) , \\
	R_2 \leq \log q- H(N_1) , \\
	R_1+R_2  \leq \log q +H(N_3) - H(N_1)- H(N_2)
\end{array}
\right.\right\}.
\Label{HH3}
\end{align}
Also, the individual secrecy achievable rate region \eqref{HH2} is evaluated to
 \begin{align}
{\cal R}_{I, N}\supseteq {\cal A}_I:=
\left\{(R_1,R_2)\left|
\begin{array}{l}
	R_1 \leq \log q- H(N_2) ,\\
	R_2 \leq \log q- H(N_1) , \\
	\max(R_1,R_2) \leq \log q +H(N_3) - H(N_1)- H(N_2)
\end{array}
\right.\right\}.
\Label{HH4}
\end{align}

We define the outer region
 \begin{align}
{\cal A}_{O}:=\left\{(R_1,R_2)\left|
\begin{array}{l}
	R_1 \leq \log q- H(N_2),\\
	R_2 \leq \log q- H(N_1)  
\end{array}
\right.\right\}.
\Label{HH4-2}
\end{align}

Considering the case when $V_i=X_i$ is the uniform random number for $i=1,2$,
we have 
the following corollary of Theorem \ref{Cor: S-MAC-KS}.
\begin{corollary}\Label{Cor:1} 
Under the channel model \eqref{AD1} -- \eqref{AD3}, 
we have the following relations.
\begin{align}
{\cal A}_J \subset {\cal C}_{J,S,N,2},\quad
{\cal A}_I \subset {\cal C}_{I,S,N,2}.\Label{NZT-AD}
\end{align}
\end{corollary} 

The reference \cite[Corollary 1]{GKYG2013} showed 
	the achievability of the rate region
	${\cal A}_{J}$ under the weak secrecy for the binary additive TW-WC, i.e., 
	\begin{align}
		{\cal A}_J \subset {\cal C}_{J,W,N,2}.
	\end{align}
	Corollary \ref{Cor:1} extended the above relation to additive TW-WC over any finite field and to the case with strong secrecy.

Next, 
to apply Lemma \ref{L3T} with the above choice of
$X_i$ and $V_i$ for $i=1,2$, 
we prepare the R\'{e}nyi entropy $H_{1+s}(X)$ of a random variable $X$
which is defined as 
$-\frac{1}{s}\log \sum\limits_{x}P_X(x)^{1+s} $. 
Then, for the additive TW-WC, 
the quantities 
$e^{-sI_{\frac{1}{1+s}}^{\uparrow} (Y_2 ;  X_{1}| X_{2})}$
and $e^{-sI_{\frac{1}{1+s}}^{\uparrow} (Y_1 ;  X_{2}| X_{1})}$
are calculated as
\begin{align}
	e^{-sI_{\frac{1}{1+s}}^{\uparrow} (Y_2 ;  X_{1}| X_{2})}=
	e^{-s\log q+ sH_{\frac{1}{1+s}}(N_2)},\quad
	e^{-sI_{\frac{1}{1+s}}^{\uparrow} (Y_1 ;  X_{2}| X_{1})}=
	e^{-s\log q+ sH_{\frac{1}{1+s}}(N_1)}. \Label{eq:ReliabilityA} 
\end{align}
Also, 
$e^{s I^{\downarrow}_{1+s} ( Z ; X_1)}$,
$e^{s I^{\downarrow}_{1+s} ( Z ; X_2 )}$, and
$e^{s I^{\downarrow}_{1+s} ( Z ; X_1,X_2)}$
are calculated as
\begin{align}
	e^{s I^{\downarrow}_{1+s} ( Z ; X_1)}=1,\quad
	e^{s I^{\downarrow}_{1+s} ( Z ; X_2 )}=1,\quad
	e^{s I^{\downarrow}_{1+s} ( Z ; X_1,X_2)}=e^{s \log q-sH_{1+s}(N_3)}.
\end{align}
Hence, Lemma \ref{L3T} implies the following corollary. 

\begin{corollary}\Label{C3T}
For 
a rate pair $(R_1,R_2)$ and two positive numbers $r_1,r_2$,
there exists a sequence of $(e^{nR_1}, e^{nR_2}, n)$ non-adaptive codes $\{\Co_n\}$ with the 2nd type decoder such that
	\begin{align} 
	  P_{e}^n(\Co_n) &\leq 2\left(e^{ns\left(R_1+r_1 -\log q+ H_{\frac{1}{1+s}}(N_2)\right)}
	  +e^{ns\left(R_2+r_2 -\log q+ H_{\frac{1}{1+s}}(N_1)\right)} \right)
, \Label{eq:ReliabilityC} \\
	  I(M_1, M_2;Z^n|\Co_n)	&\leq	
	  2\left(e^{ns\left(\log q -H_{1+s}(N_3)-r_1-r_2\right)}+e^{-nsr_1}+e^{-nsr_2}\right).
\Label{eq:SSecC}
\end{align}
Also, there exists a sequence of $(e^{nR_1}, e^{nR_2}, n)$ non-adaptive codes $\{\Co_n\}$ such that
	\begin{align} 
	  P_{e}^n(\Co_n) &\leq 
	 3\left(e^{ns\left(R_1+r_1 -\log q+ H_{\frac{1}{1+s}}(N_2)\right)}
	  +e^{ns\left(R_2+r_2 -\log q+ H_{\frac{1}{1+s}}(N_1)\right)} \right)
 , \Label{eq:Reliability2C} \\
	  I(M_1;Z^n|\Co_n)	&\leq 
	  	  3\left(e^{ns\left(\log q -H_{1+s}(N_3)-r_1-r_2-R_2\right)}+e^{-nsr_1}+e^{-ns(r_2+R_2)}\right)
, \Label{eq:SSec2C} \\
	  I( M_2;Z^n|\Co_n)	&	\leq 3\left(e^{ns\left(\log q -H_{1+s}(N_3)-r_1-r_2-R_1\right)}+e^{-nsr_2}+e^{-ns(r_1+R_1)}\right)
	  . \Label{eq:SSec3C}
\end{align}
\end{corollary}

\begin{theorem}\Label{add-F} 
When 
there exists an
independent random variable $N_1'$ or $N_2'$ from $N_1$ and $N_2$
such that
$Z=a_3a_1^{-1} Y_1+N_1'$ or
$Z=a_3a_2^{-1} Y_2+N_2'$, 
the equalities in \eqref{HH3} and \eqref{HH4} hold, i.e., ${\cal R}_{J, N}= {\cal A}_J$ and ${\cal R}_{I, N} = {\cal A}_I.$
\end{theorem}
\begin{IEEEproof}
	See Appendix \ref{sec: proof of Theorem add-F} for a detailed proof.
\end{IEEEproof}

Notice that 
the condition $Z=a_3a_1^{-1} Y_1+N_1'$
is a stronger condition than the condition
$a_3a_1^{-1}=b_3b_1^{-1} $.
Also,
the condition 
$Z=a_3a_2^{-1} Y_2+N_2'$ is a stronger condition than the condition
$a_3a_2^{-1}=b_3b_2^{-1} $.

\begin{theorem}\Label{add-C} 
We have the following relation.
 \begin{align}
{\cal C}_{I,W,A,1} &\subseteq {\cal A}_{O}
\Label{HH4-7B}
\end{align}
When 
there exists an
independent random variable $N_3'$ from $N_1$ and $N_2$
such that $N_3=
\frac{b_3}{b_1} {N}_1+  \frac{a_3}{a_2}{N}_2+N_3'$,
we have
\begin{align}
{\cal C}_{J,W,N,1}
\subseteq {\cal A}_{J}.
\Label{HH4-6}
\end{align}
Since ${\cal C}_{J,W,N,2}\subset  {\cal C}_{J,W,N,1}$,
with Corollary \ref{Cor:1}, 
the relation \eqref{HH4-6} implies the relation
\begin{align}
	{\cal C}_{J,b,N,d} =  {\cal R}_{J, N}=  {\cal A}_{J}
	\Label{HH8-6}
\end{align}
for $b=S,W$ and $d=1,2$.
\end{theorem}
\begin{IEEEproof}
	See Appendix \ref{sec: proof of Theorem add-C} for a detailed proof.
\end{IEEEproof}

We notice that Theorem \ref{add-C} implies the equality in \eqref{HH3}
with a condition different from Theorem \ref{add-F}.

When $\max\{H(N_1), H(N_2)\} \le H(N_3)$, we have
$H(N_1)+H(N_2)-H(N_3) \le \min\{H(N_1), H(N_2)\}$, which indicates that the first and second condition in \eqref{HH4} implies the third condition in \eqref{HH4}. Hence, under this condition, the region ${\cal A}_I$ is simplified to ${\cal A}_{O}$.
Hence, we have the following corollary.

\begin{corollary}\Label{add-C2} 
When $\max\{H(N_1), H(N_2)\}\le H(N_3)$, 
we have the following relation.
 \begin{align}
{\cal C}_{I,b,c,d}={\cal A}_{O}
\Label{HH4-7}
\end{align}
for $b=S,W$, $c=N,A$, and $d=1,2$.
\end{corollary}

\section{Gaussian two-way wiretap channel}\label{Sec: Gaussion TW-WC}
In Gaussian two-way wiretap channel, 
$X_1,X_2,Y_1,Y_2,Z$ are variables in $\RR$.
Then, we have
\begin{align}
Y_1&= a_1 X_1+b_1 X_2+N_1, \Label{AG1}\\
Y_2&= a_2 X_1+ b_2 X_2+N_2, \Label{AG2}\\
Z&=  a_3 X_1+b_3 X_2+N_3,\Label{AG3}
\end{align}
where $N_1$, $N_2$, $N_3$ are 
Gaussian random variables with 
mean 0 and variance $v_1,$ $v_2,$ $v_3$.
Also, $a_1,b_1,a_2,b_2,a_3,b_3$ are constants.
Choosing $g_i$ as $g_i(X_i):=X_i^2$, we consider the cost constraint.

Assume that $X_1$($X_2$) is a Gaussian random variable with mean $0$ and variance $p_1 $
($p_2$).
Then, we have
\begin{align}
I(Y_2;X_1|X_2)&=
\frac{1}{2}(\log (a_2^2 p_1 + v_2)-\log v_2  ) \\
I(Y_1;X_2|X_1)&=
\frac{1}{2}(\log (b_1^2 p_2 + v_1)-\log v_1  ) \\
I(Z;X_1)&=\frac{1}{2}(\log (a_3^2 p_1+ b_3^2 p_2 + v_3)-\log ( b_3^2 p_2 + v_3)  ) \\
I(Z;X_2)&= \frac{1}{2}(\log (a_3^2 p_1+ b_3^2 p_2 + v_3)-\log ( a_3^2 p_1 + v_3)  )\\
I(X_1, X_2;Z)&=
\frac{1}{2}(\log (a_3^2 p_1+ b_3^2 p_2 + v_3)-\log v_3  ) .
\end{align}
Therefore, the joint secrecy achievable rate region \eqref{BPW1} is converted to
 \begin{align}
{\cal R}_J^{p_1,p_2} := 
\left\{(R_1,R_2)\left|
\begin{array}{l}
	R_1 \leq \frac{1}{2} \Big(
\log  (a_2^2 p_1+v_2)
- \log   v_2
-\log (a_3^2 p_1+ b_3^2 p_2+v_3) 
+\log (b_3^2 p_2+v_3)  \Big)\\
	R_2 \leq  
	\frac{1}{2} \Big(
\log  (b_1^2 p_2+v_1)
- \log  v_1 
-\log (a_3^2 p_1+ b_3^2 p_2+v_3) 
+\log (a_3^2 p_1+v_3)  \Big)\\
	R_1+R_2  \leq  
	\frac{1}{2} \Big(
\log (b_1^2 p_2+v_1)
+\log  (a_2^2 p_1+v_2)
-\log (a_3^2 p_1+ b_3^2 p_2+v_3) \\
\hspace{12ex}+\log v_3 -\log  v_1- \log v_2) \Big).
\end{array}
\right.\right\}.
\Label{HH5}
\end{align}
An inner bound on the strong joint secrecy capacity region is obtained as 
\begin{equation}
	{\cal A}_J^{c_1,c_2} := cl. \bigcup_{
			0\leq p_1\leq c_1; 
			0\leq p_2\leq c_2}  {\cal R}_J^{p_1,p_2}.
\end{equation}
Also, the individual secrecy achievable rate region \eqref{BPW2} is converted to
 \begin{align}
{\cal R}_I^{p_1,p_2}:=
\left\{(R_1,R_2)\left|
\begin{array}{l}
	R_1 \leq \frac{1}{2} \Big(
\log  (a_2^2 p_1+v_2)
- \log   v_2
-\log (a_3^2 p_1+ b_3^2 p_2+v_3) 
+\log (b_3^2 p_2+v_3)  \Big)\\
	R_2 \leq  
	\frac{1}{2} \Big(
\log  (b_1^2 p_2+v_1)
- \log  v_1 
-\log (a_3^2 p_1+ b_3^2 p_2+v_3) 
+\log (a_3^2 p_1+v_3)  \Big)\\
	\max(R_1,R_2) \leq 
	\frac{1}{2} \Big(
\log (b_1^2 p_2+v_1)
+\log  (a_2^2 p_1+v_2)
-\log (a_3^2 p_1+ b_3^2 p_2+v_3) \\
\hspace{16ex}+\log v_3 -\log  v_1- \log v_2) \Big).
\end{array}
\right.\right\}.
\Label{HH6}
\end{align}
An inner bound on the strong individual secrecy capacity region is obtained as 
\begin{equation}
	{\cal A}_I^{c_1,c_2} := cl. \bigcup_{0\leq p_1\leq c_1; 0\leq p_2\leq c_2}  {\cal R}_I^{p_1,p_2}.
\end{equation}

For the direct part, we have the following theorem.
\begin{theorem}\Label{DG-C}
we have the following relations.
\begin{align}
{\cal A}_J^{c_1,c_2} \subset {\cal C}_{J,S,N,2}^{P,c_1,c_2},\quad
{\cal A}_I^{c_1,c_2} \subset {\cal C}_{I,S,N,2}^{P, c_1,c_2}.\Label{NZT-CG}
\end{align}
\end{theorem}

As observed in \cite{PB2011} that the reference \cite{TY2007, TY2008} has showed the achievability of the rate region
${\cal A}_J^{c_1,c_2}$ under the weak secrecy and the average cost constraint,  i.e., 
\begin{align}
{\cal A}_J^{c_1,c_2} \subset {\cal C}_{J,W,N,2}^{A,c_1,c_2}.
\end{align} 
Theorem \ref{DG-C} extended the above relation to the case with strong secrecy and 
the peak cost constraint.
It is worth mentioning that this region (under the weak secrecy and the average cost constraint) is extended in \cite{GKYG2013} by applying adaptive coding. Here, we restrict our focus to the non-adaptive coding.

For the converse part, we have the following theorem.
\begin{theorem}\Label{G-C} 
When $ \frac{b_3^2}{b_1^2}v_1+ \frac{a_3^2}{a_2^2}v_2\le v_3$,
we have the following relations.
 \begin{align}
{\cal C}_{J,W,N,1}^{A,c_1,c_2} & \subset 
{\cal A}_{J, N}^{c_1,c_2}:=
\left\{(R_1,R_2)\left|
\begin{array}{l}
	R_1 \le \frac{1}{2} \Big(
	\log  (a_2^2 c_1+v_2)
	- \log   v_2 \Big)\\
	R_2 \leq  
	\frac{1}{2} \Big(
	\log  (b_1^2 c_2+v_1)
	- \log  v_1 \Big)\\
	R_1+R_2 \le
	\frac{1}{2} \Big(
	\log (b_1^2 c_2+v_1)
	+\log (a_2^2 c_1+v_2)
	-\log (a_3^2 c_1+ b_3^2 c_2+v_3) \\
\hspace{12ex}	+\log v_3 -\log v_1- \log v_2) \Big)
\end{array}
\right.\right\}.
\Label{HH34}
\end{align}
\end{theorem}
\begin{IEEEproof}
	See Appendix \ref{sec: proof of Theorem G-C} for a detailed proof.
\end{IEEEproof}

\section{Conclusion}\label{Sec: Conclusion}
In this paper, we have studied the TW-WC under a strong joint or individual secrecy constraint. 
Restricting our focus to non-adaptive coding, we have derived inner bounds on the strong secrecy capacity regions for the TW-WC and characterized the secrecy and error exponents by the {\it conditional R\'enyi mutual information}. 
Moreover, we have introduced the peak cost constraint and extended our results by using the constant composition codes, where both the secrecy and error exponents are accordingly characterized by {\it a modification of R\'enyi mutual information}. 
Our method works even when a pre-noisy processing is employed based on a conditional distribution in the encoder and the results are valid for the general discrete memoryless TW-WC without/with cost constraint.
In addition, we have applied our results to the case with additive noise channel over a finite field
and the case with the Gaussian noise.
In these examples,
we have derived outer bounds in various settings
in addition to inner bounds.

Our results can naturally be extended in several directions.
First, our results can be generalized to other multi-user communication channels. For the TW-WC, we did not include the scenario with one-sided secrecy, since the analysis is straightforward by following our approaches for the scenarios with joint/individual secrecy. Besides, it is not difficult to generalize our results to channels with more than two legitimate users under diverse secrecy constraints. In fact, our technique for strong secrecy analysis
reflects the channel resolvability for the $k$-transmitter MAC \cite{HC2019}; our error exponent could be regarded as a generalization of Gallager's error exponent \cite{Gallager68}; and our method covers also the channels with cost constraints by using random coding on constant composition codes, all together laying a solid foundation for yielding inner bounds and the corresponding secrecy and error exponents for various settings of multi-user communication channels with strong secrecy constraint that could be requested differently depending on the actual applications. 
Besides, in this paper we have focused on the non-adaptive codes only. As already witnessed in \cite{GKYG2013} for the Gaussian TW-WC under joint weak secrecy, the adaptive coding could enlarge the inner bounds that were established based on the non-adaptive coding. A further notice is that the adaptive coding scheme proposed in \cite{GKYG2013} could be extended to the scenario with joint strong secrecy as shown in \cite{PB2011}. Clearly, this extension is not limited to the joint secrecy only. It can be extended to the TW-WCs with individual secrecy or one-sided secrecy as well, to potentially improve the existing inner bounds, while the analysis for secrecy and reliability needs to be revised since the input is adapted to the previously received signals in the adaptive coding. A more challenging task could be to find an adaptive coding scheme, differently from \cite{GKYG2013}, that actually could improve the secrecy sum-rate that is achieved by non-adaptive coding; or to provide th
 e converse proof if adaption is useless to improve the secrecy sum-rate.   

\section*{Acknowledgments}
MH is very grateful to Prof. Te Sun Han  
for explaining the relation between 
the references \cite{Han,Sreekumar}.

\appendices

\section{Proof of Lemma \ref{LL1}}\label{App: proof of Lemma LL1}
We show the closedness of ${\cal C}_{J,S(W), A(N),1(2)}^{P(A),c_1,c_2}$.
It is sufficient to show the following.
When we choose any sequence $\{\kappa_n\} $ in ${\cal C}_{J,S(W), A(N),1(2)}^{P(A),c_1,c_2} $
such that $\kappa_m \to \kappa_0$ as $m \to \infty$,
$\kappa_0$ belongs to ${\cal C}_{J,S(W), A(N),1(2)}^{P(A),c_1,c_2}$.
We choose a sequence codes $\{C_{n,m}\}_{n=1}^{\infty}$
such that $\kappa(C_{n,m}) \to \kappa_{m}$,
$P_{e}^n(\Co_{n,m}) \to 0$, and
$I(M_1, M_2;Z^n|\Co_{n,m})\to 0$
as $n \to \infty$.
For any positive integer $l>0$, we choose 
$m_l$ such that $ \|\kappa_m- \kappa_0\|\le \frac{1}{l}$.
We choose $n_l$ satisfying the following conditions;
$n_{l+1} > n_l$.
For any $n \ge n_l$,we have
$\|\kappa(C_{n,m_l}) - \kappa_{m_l}\| \le \frac{1}{l}$,
$P_{e}^n(\Co_{n,m_l}) \le \frac{1}{l}$, and
$I(M_1, M_2;Z^n|\Co_{n,m_l})\le \frac{1}{l}$.
We choose $C_n$ as 
$\Co_{n,m_l}$ for $n_l \le n < n_{l+1}$.
Then, we have 
$\kappa(C_{n}) \to \kappa_{0}$,
$P_{e}^n(\Co_{n}) \to 0$, and
$I(M_1, M_2;Z^n|\Co_{n})\to 0$ as $n \to \infty$.
Therefore, $\kappa_0$ belongs to ${\cal C}_{J,S(W), A(N),1(2)}^{I(II),c_1,c_2}$.

The closedness of ${\cal C}_{I,S(W), A(N),1(2)}^{P(A),c_1,c_2}$
can be shown in the same way.
\section{Proof of Lemma \ref{L2}}\Label{A-E}
\allowdisplaybreaks
To prove Lemma \ref{L2}, we consider
	\begin{align}
	& \mathbb{E}_{\cC} 
	\left[ e^{s D_{1+s}(  P_{Z| \cC}\| P_{Z})}\right]\nonumber\\
	\stackrel{(a)}{=} & \sum_{z} 
	\mathbb{E}_{\cC}\left[ 
	\left(\frac{1}{\sL_1 \sL_2} \sum_{j_1', j_2'} 
	P_{Z| X_1=X_{1,j_1'},X_2=X_{2,j_2'}  }(z)\right)
	\left(\frac{1}{\sL_1 \sL_2} \sum_{j_1,j_2} 
	P_{Z| X_1=X_{1,j_1}, X_2=X_{2,j_2}  }(z)\right)^{s}
	P_Z(z)^{-s} \right]\nonumber\\
	\stackrel{(b)}{=} & \sum_{z} 
	\mathbb{E}_{\cC}\left[
	\frac{1}{\sL_1 \sL_2} \sum_{j_1', j_2'} 
	P_{Z| X_1=X_{1,j_1'}, X_2=X_{2,j_2'}  }(z)\right.
	\nonumber\\
	&\left.\cdot 
	\mathbb{E}_{\cC|X_1=X_{1,j_1'}, X_2=X_{2,j_2'}  }
	\left[\left(\frac{1}{\sL_1 \sL_2} \sum_{j_1,j_2} 
	P_{Z| X_1=X_{1,j_1}, X_2=X_{2,j_2}  }(z)\right)^{s}\right]
	P_Z(z)^{-s} \right]\nonumber\\
	\stackrel{(c)}{\le} 
      & \sum_{z} 
	\mathbb{E}_{\cC}\left[
	\frac{1}{\sL_1 \sL_2} \sum_{j_1', j_2'} 
	P_{Z| X_1=X_{1,j_1'}, X_2=X_{2,j_2'}  }(z) \right.\nonumber\\
	&\left.\cdot
	\left(
	\mathbb{E}_{\cC|X_1=X_{1,j_1'}, X_2=X_{2,j_2'}  }
	\left[\frac{1}{\sL_1 \sL_2} \sum_{j_1,j_2} 
	P_{Z| X_1=X_{1,j_1}, X_2=X_{2,j_2}  }(z)\right]\right)^{s}
	P_Z(z)^{-s} \right]\nonumber\\
	=& \sum_{z} 
	\mathbb{E}_{\cC}\left[
	\frac{1}{\sL_1 \sL_2} \sum_{j_1',j_2'} 
	P_{Z| X_1=X_{1,j_1'},X_2=X_{2,j_2'}  }(z) \right.\nonumber\\
	& \left. \cdot \left(
	\mathbb{E}_{\cC|X_1=X_{1,j_1'},X_2=X_{2,j_2'}  }
	\left[ \frac{1}{\sL_1 \sL_2} 
	\sum_{\mathcal{S} \subseteq  \{1,2\}}
	\sum_{\scriptsize 
		\begin{array}{c}
		j_i \neq j_i'\\
		i \in \mathcal{S}
		\end{array}
	}
	P_{Z| (X_i=X_{i,j_i})_{i \in \mathcal{S}} ,
		(X_i=X_{i,j_i'})_{i \in {\mathcal{S}}^c}   }(z)\right]\right)^{s}
	P_Z(z)^{-s} \right]\nonumber\\
	=& \sum_{z} 
	\mathbb{E}_{\cC} \left[
	\frac{1}{\sL_1 \sL_2} \sum_{j_1', j_2'} 
	P_{Z| X_1=X_{1,j_1'}, X_2=X_{2,j_2'}  }(z)
	\left(
	\frac{1}{\sL_1 \sL_2} 
	\sum_{\mathcal{S} \subseteq  \{1,2\}}
	\left(\prod_{i \in \mathcal{S}}(\sL_i-1)\right)
	P_{Z| (X_i=X_{i,j_i'})_{i \in {\mathcal{S}}^c}   }(z)\right)^{s}
	P_Z(z)^{-s} \right]\nonumber\\
	\stackrel{(d)}{\le} & \sum_{z} 
	\mathbb{E}_{\cC}\left[
	\frac{1}{\sL_1 \sL_2} \sum_{j_1', j_2'} 
	P_{Z| X_1=X_{1,j_1'}, X_2=X_{2,j_2'}  }(z)
	\left(
	\frac{1}{\sL_1^s \sL_2^s} 
	\sum_{\mathcal{S} \subseteq  \{1,2\}}
	\left(\prod_{i \in {\mathcal{S}}}(\sL_i-1)^s\right)
	P_{Z| (X_i=X_{i,j_i'})_{i \in {\mathcal{S}}^c}   }(z)^{s}\right)
	P_Z(z)^{-s} \right]\nonumber\\
	\stackrel{(e)}{\le} & \sum_{z} 
	\mathbb{E}_{\cC}\left[
	\frac{1}{\sL_1 \sL_2} \sum_{j_1', j_2'} 
	P_{Z| X_1=X_{1,j_1'}, X_2=X_{2,j_2'}  }(z)
	\left(
	\sum_{\mathcal{S} \subseteq  \{1,2\}}
	\frac{1}{\prod_{i \in {\mathcal{S}}^c}\sL_i ^s} 
	P_{Z| (X_i=X_{i,j_i'})_{i \in {\mathcal{S}}^c}   }(z)^{s}\right)
	P_Z(z)^{-s} \right]\nonumber\\
	= & \sum_{z} 
	\frac{1}{\sL_1 \sL_2} \sum_{j_1', j_2'} 
	\left(
	\sum_{{\mathcal{S}} \subseteq \{1,2\}}
	\frac{1}{\prod_{i \in {\mathcal{S}}^c}\sL_i ^s} 
	\mathbb{E}_{\cC}\left[
	P_{Z| X_1=X_{1,j_1'},X_2=X_{2,j_2'}  }(z)
	P_{Z| (X_i=X_{i,j_i'})_{i \in {\mathcal{S}}^c}   }(z)^{s} P_Z(z)^{-s}\right]\right)
	\nonumber\\
	= & 
	\frac{1}{\sL_1 \sL_2} \sum_{j_1', j_2'} 
	\sum_{{\mathcal{S}} \subseteq \{1,2\}}
	\frac{1}{\prod_{i \in {\mathcal{S}}^c}\sL_i^s } 
	\sum_{ (x_i)_{i \in {\mathcal{S}}^c} }
	\left(\prod_{i \in {\mathcal{S}}^c} P_{X_i}( x_{i} )\right)
	\sum_{z} 
	P_{Z| (X_i=x_{i})_{i \in {\mathcal{S}}^c}   }(z)^{1+s} P_Z(z)^{-s} \nonumber\\
	\stackrel{(f)}{=} & 
	\sum_{\mathcal{S}\subseteq \{1,2\}}
	\frac{1}{ \prod_{i \in \mathcal{S}^c} \sL_i^s} 
	e^{s I_{1+s}^{\downarrow} ( Z ; X_{\mathcal{S}^c}) } 
	= \sum_{\mathcal{S}\subseteq  \{1,2\}}
	\frac{1}{ \prod_{i \in \mathcal{S}} \sL_i^s} 
	e^{s I_{1+s}^{\downarrow} ( Z ; X_{\mathcal{S}}) } 
      \nonumber\\
	= & 
	1+\sum_{\mathcal{S}\neq \emptyset, \mathcal{S}\subseteq  \{1,2\}}
	\frac{1}{ \prod_{i \in \mathcal{S}} \sL_i^s} 
	e^{s I_{1+s}^{\downarrow} ( Z ; X_{\mathcal{S}}) },\Label{lem1: step f}
	\end{align}
where $X_{\mathcal{S}}=\{X_i|i\in \mathcal{S}\}$ with $P_{X_{\mathcal{S}}}=\prod_{i\in \mathcal{S}}P_{X_i}.$
In more details, $(a)$ is by the definition of R\'enyi relative entropy in \eqref{eqn: Renyi relative entropy}; 
$(b)$ is by the law of total expectation; 
$(c)$ is by the concavity of $x^s$ for $0\leq s\leq 1$; 
$(d)$ is due to  the fact that $(\sum_i a_i)^s\leq \sum\limits_i a_i^s$ for $a_i\geq 0$ and $0\leq s\leq 1$; 
$(e)$ follows from $\sL_i-1 \le \sL_i$;
and $(f)$ is by the definition of R\'enyi mutual information in \eqref{Ne1}. Taking the logarithm to \eqref{lem1: step f}, we have
\begin{equation}
	s \mathbb{E}_{\cC} \left[ 
	D_{1+s}(  P_{Z| \cC}\| P_{Z}) \right]
	\le  
	\log \Big( 1+\sum_{\mathcal{S}\neq \emptyset, \mathcal{S}\subseteq  \{1,2\}}
	\frac{1}{ \prod_{i \in \mathcal{S}} \sL_i^s} 
	e^{s I_{1+s}^{\downarrow} ( Z ;  X_{\mathcal{S}}) }
	\Big) 
	\le 
	\sum_{{\mathcal{S}}\neq \emptyset, \mathcal{S}\subseteq \{1,2\}}
	\frac{1}{ \prod_{i \in \mathcal{S}} \sL_i^s} 
	e^{s I_{1+s}^{\downarrow} ( Z ; X_{\mathcal{S}}) }.
\end{equation}

\section{Fourier-Motzkin elimination to derive individual secrecy region}\Label{AP1}
To derive individual secrecy region, recall that we have the following rate constraints:
\begin{align}
	r_1 &> I(Z;X_1);\Label{C1}\\
	r_2 &> I(Z;X_2);\Label{C2}\\
	r_1+r_2+\min(R_1,R_2) &> I(Z;X_1, X_2);\Label{C3}\\
	R_1+r_1 &< I(Y_2;X_1|X_2);\Label{C4}\\
	R_2+r_2 &< I(Y_1;X_2|X_1).\Label{C5}
\end{align}
First consider \eqref{C1}, \eqref{C3} and \eqref{C4} to remove $r_1.$ We obtain
\begin{align}
	R_1 &< I(Y_2;X_1|X_2)-I(Z;X_1);\Label{C6}\\
	r_2 &> R_1-I(Y_2;X_1|X_2)+I(Z;X_1, X_2)-\min(R_1,R_2).\Label{C7}
\end{align}
Now consider \eqref{C2}, \eqref{C5} and \eqref{C7} to remove $r_2,$ We obtain
\begin{align}
	R_2 &\leq I(Y_1;X_2|X_1)-I(Z;X_2);\Label{C8}\\
	\max\{R_1, R_2\} & \leq  I(Y_2;X_1|X_2) + I(Y_1;X_2|X_1) - I(X_1, X_2;Z).\Label{C9}
\end{align} 
The achievable region is thus established by \eqref{C6},\eqref{C8} and \eqref{C9}.

\section{Proof of Lemma \ref{L2C}}\Label{A-F}
\allowdisplaybreaks

Using \eqref{eq3-X}, we have
\begin{align}
\breve{P}_{X_i^n | V_{i}^n} \le \nu_n(|{\cal X}_i|)   P_{X_i|V_i}^{n} ,\quad
\breve{P}_{V_{i}^n} \le  \nu_n(|{\cal V}_i|)   P_{V_i}^{n} .\Label{XNA}
\end{align}

Thus, we have
\begin{align}
P_{Z|X_1X_2}^n (z^n|x_1^n, x_2^n) \breve{P}_{X^n_{2}} (x_2^n) 
	\breve{P}_{X^n_{1}|V^n_{1}}(x_1^n|v_1^n)
\le \nu_n(|{\cal X}_1|) \nu_n(|{\cal X}_2|)
P_{Z|X_1X_2}^n (z^n|x_1^n, x_2^n)
 {P}_{X_2}^n  (x_2^n) 
 {P}_{X_{1}|V_{1}}^n (x_1^n|v_1^n).
\end{align}
Taking the sum for $x_1^n,x_2^n$, we have
\begin{align}
P_{Z|X_1X_2}^n 
	\cdot \breve{P}_{X^n_{2}}  
	\cdot 	\breve{P}_{X^n_{1}|V^n_{1}}
\le \nu_n(|{\cal X}_1|)\nu_n(|{\cal X}_2|)
P_{Z|X_1X_2}^n 
	\cdot {P}_{X_{2}}^n  
	\cdot {P}_{X_{1}|V_{1}}^n.\Label{MLD}
\end{align}

Thus, we have
\begin{align}
&P_{Z|X_1X_2}^n (z^n|x_1^n, x_2^n) 
 	\breve{P}_{X^n_{\{1,2\}}|V^n_{\{1,2\}}}(x_{\{1,2\}}^n|v_{\{1,2\}}^n) 
=
P_{Z|X_1X_2}^n (z^n|x_1^n, x_2^n) 
 	\breve{P}_{X^n_{1}|V^n_{1}}(x_1^n|v_1^n)
 	\breve{P}_{X^n_{2}|V^n_{2}}(x_2^n|v_2^n)  \nonumber\\
\le & \nu_n(|{\cal X}_1|) \nu_n(|{\cal X}_2|)
P_{Z|X_1X_2}^n (z^n|x_1^n, x_2^n)
{P}_{X_{1}|V_{1}}^n (x_1^n|v_1^n)
{P}_{X_{2}|V_{2}}^n (x_2^n|v_2^n).
\end{align}
Taking the sum for $x_1^n,x_2^n$, we have
\begin{align}
P_{Z|X_1X_2}^n 
	\cdot 	\breve{P}_{X^n_{\{1,2\}}|V^n_{\{1,2\}}}
\le \nu_n(|{\cal X}_1|)\nu_n(|{\cal X}_2|)
P_{Z|X_1X_2}^n 
	\cdot {P}_{X_{\{1,2\}}|V_{\{1,2\}}}^n.
\end{align}
In the same way, we have
\begin{align}
P_{Z|X_1X_2}^n 
	\cdot 	\breve{P}_{X^n_{\{1,2\}}}
	\le \nu_n(|{\cal X}_1|)\nu_n(|{\cal X}_2|)
P_{Z|X_1X_2}^n 
	\cdot {P}_{X_{\{1,2\}}}^n.
\end{align}
In summary, we have
\begin{align}
P_{Z|X_1X_2}^n 
	\cdot \breve{P}_{X^n_{\mathcal{S}^c}}  
	\cdot 	\breve{P}_{X^n_{\mathcal{S}}|V^n_{\mathcal{S}}}
\le 
\nu_n(|{\cal X}_1|)\nu_n(|{\cal X}_2|)
P_{Z|X_1X_2}^n 
	\cdot {P}_{X_{\mathcal{S}^c}}^n  
	\cdot {P}_{X_{\mathcal{S}}|V_{\mathcal{S}}}^n.
\end{align}

Using \eqref{MVK}, we have
\begin{align}
	e^{s I_{\frac{1}{1-s}}^{\uparrow} ( Z^n ; V^n_{\mathcal{S}})[P_{Z|X_1X_2}^n 
	\cdot \breve{P}_{X^n_{\mathcal{S}^c}}  
	\cdot 	\breve{P}_{X^n_{\mathcal{S}}|V^n_{\mathcal{S}}} 
	\times \breve{P}_{V^n_{\mathcal{S}}}]}
	\le 
\nu_n(|{\cal X}_1|)\nu_n(|{\cal X}_2|)
	e^{s I_{\frac{1}{1-s}}^{\uparrow} ( Z^n ; V^n_{\mathcal{S}})[P_{Z|X_1X_2}^n 
	\cdot {P}_{X_{\mathcal{S}^c}}^n  
	\cdot {P}_{X_{\mathcal{S}}|V_{\mathcal{S}}}^n 
	\times \breve{P}_{V^n_{\mathcal{S}}}]}\Label{ZLP}
\end{align}

Therefore, for $s \in [0,1]$, we have
	\begin{align}
	& \mathbb{E}_{\cC} \left[
	e^{s D_{1+s}(  P_{Z^n| \cC}\| P_{Z^n})}\right]-1 \nonumber\\
	\stackrel{(a)}{\le} & 
	\sum_{\mathcal{S}\neq \emptyset, \mathcal{S}\subseteq  \{1,2\}}
	\frac{1}{ \prod_{i \in \mathcal{S}} \sL_i^s} 
	e^{s I_{1+s}^{\downarrow} ( Z^n ; V^n_{\mathcal{S}})[P_{Z|X_1X_2}^n 
	\cdot \breve{P}_{X^n_{\mathcal{S}^c}}  
	\cdot 	\breve{P}_{X^n_{\mathcal{S}}|V^n_{\mathcal{S}}} 
	\times \breve{P}_{V^n_{\mathcal{S}}}
	] } \nonumber\\
	\stackrel{(b)}{\le} &
	\sum_{\mathcal{S}\neq \emptyset, \mathcal{S}\subseteq  \{1,2\}}
	\frac{1}{ \prod_{i \in \mathcal{S}} \sL_i^s} 
	e^{s I_{\frac{1}{1-s}}^{\uparrow} ( Z^n ; V^n_{\mathcal{S}})[P_{Z|X_1X_2}^n 
	\cdot \breve{P}_{X^n_{\mathcal{S}^c}}  
	\cdot 	\breve{P}_{X^n_{\mathcal{S}}|V^n_{\mathcal{S}}} 
	\times \breve{P}_{V^n_{\mathcal{S}}}
	] } \nonumber\\
	\stackrel{(c)}{\le} &
	\sum_{\mathcal{S}\neq \emptyset, \mathcal{S}\subseteq  \{1,2\}}
	\frac{\nu_n(|{\cal X}_1|)\nu_n(|{\cal X}_2|)
}{ \prod_{i \in \mathcal{S}} \sL_i^s} 
	e^{s I_{\frac{1}{1-s}}^{\uparrow} ( Z^n ; V^n_{\mathcal{S}})[P_{Z|X_1X_2}^n 
	\cdot {P}_{X_{\mathcal{S}^c}}^n  
	\cdot {P}_{X_{\mathcal{S}}|V_{\mathcal{S}}}^n 
	\times \breve{P}_{V^n_{\mathcal{S}}}
	] } \nonumber\\
	= &
	\sum_{\mathcal{S}\neq \emptyset, \mathcal{S}\subseteq  \{1,2\}}
	\frac{\nu_n(|{\cal X}_1|)\nu_n(|{\cal X}_2|)
}{ \prod_{i \in \mathcal{S}} \sL_i^s} 
	e^{s I_{\frac{1}{1-s}}^{\uparrow} ( Z^n ; V^n_{\mathcal{S}})[P_{Z|V_{\mathcal{S}}}^n 
	\times \breve{P}_{V^n_{\mathcal{S}}}
	] } \nonumber\\
	\stackrel{(d)}{=} &
	\sum_{\mathcal{S}\neq \emptyset, \mathcal{S}\subseteq  \{1,2\}}
	\frac{\nu_n(|{\cal X}_1|)\nu_n(|{\cal X}_2|)
}{ \prod_{i \in \mathcal{S}} \sL_i^s} 
	e^{s \min_{Q_{Z^n}} D_{\frac{1}{1-s}} ( 
	P_{Z|V_{\mathcal{S}}}^n \times \breve{P}_{V^n_{\mathcal{S}}}\| Q_{Z^n }\times \breve{P}_{V^n_{\mathcal{S}}}
	) } \nonumber\\
	\stackrel{(e)}{\le} &
	\sum_{\mathcal{S}\neq \emptyset, \mathcal{S}\subseteq \{1,2\}}
	\frac{\nu_n(|{\cal X}_1|)\nu_n(|{\cal X}_2|)
}{ \prod_{i \in \mathcal{S}} \sL_i^s} 
	e^{s \min_{Q_{Z}} D_{\frac{1}{1-s}} ( 
	P_{Z|V_{\mathcal{S}}}^n \times \breve{P}_{V^n_{\mathcal{S}}}\| Q_Z^n\times \breve{P}_{V^n_{\mathcal{S}}}
	) } \nonumber\\
	= &
	\sum_{\mathcal{S}\neq \emptyset, \mathcal{S}\subseteq  \{1,2\}}
	\frac{\nu_n(|{\cal X}_1|)\nu_n(|{\cal X}_2|)
}{ \prod_{i \in \mathcal{S}} \sL_i^s} 
	\min_{Q_{Z}} 
      \sum_{v^n_{\mathcal{S}} \in T_{P_{V_{\mathcal{S}}}}^n } \breve{P}_{V^n_{\mathcal{S}}}(v^n_{\mathcal{S}})
      e^{s D_{\frac{1}{1-s}} ( P_{Z|V_{\mathcal{S}}}^n (\cdot| v_{\mathcal{S}}^n)  \|  Q_{Z}^n )} \nonumber \\
	\stackrel{(f)}{=}&
	\sum_{\mathcal{S}\neq \emptyset, \mathcal{S}\subseteq  \{1,2\}}
	\frac{\nu_n(|{\cal X}_1|)\nu_n(|{\cal X}_2|)
}{ \prod_{i \in \mathcal{S}} \sL_i^s} 
	\min_{Q_{Z}}
      e^{  n s \sum_{v_{\mathcal{S}}}
      P_{V_{\mathcal{S}}} (v_{\mathcal{S}})  
      D_{\frac{1}{1-s}} ( P_{Z|V_{\mathcal{S}}=v_{\mathcal{S}}}  \|  Q_{Z} )} 
       \nonumber\\
	= &
	\sum_{\mathcal{S}\neq \emptyset, \mathcal{S}\subseteq  \{1,2\}}
	\frac{\nu_n(|{\cal X}_1|)\nu_n(|{\cal X}_2|)
}{ \prod_{i \in \mathcal{S}} \sL_i^s} 
      e^{  n s \min_{Q_{Z}}  \sum_{v_{\mathcal{S}}}
      P_{V_{\mathcal{S}}} (v_{\mathcal{S}})  
      D_{\frac{1}{1-s}} ( P_{Z|V_{\mathcal{S}}=v_{\mathcal{S}}}  \|  Q_{Z} )}
       \nonumber \\
=&	  \sum_{\mathcal{S}\neq \emptyset, \mathcal{S} \subset  \{1,2\}}
	\frac{\nu_n(|{\cal X}_1|)\nu_n(|{\cal X}_2|)}{ \prod_{i \in \mathcal{S}} \sL_i^s} 
e^{n s \breve{I}_{\frac{1}{1-s}}^{\uparrow} ( Z ; V_{\mathcal{S}})  },
\end{align}
where steps $(a)$, $(b)$, and $(c)$,
follow from 
\eqref{lem1: step f} in the proof of Lemma \ref{L2}, 
\eqref{MMK}, and
\eqref{ZLP}, respectively.
Also, 
 $(d)$ and $(e)$ 
follow from \eqref{Ne3} and 
the restriction of the range of the minimization from $ Q_{Z^n}$
to $Q_Z^n$, respectively.
Also, $(f)$ follows from the equations 
\begin{align}
D_{\frac{1}{1-s}} ( P_{Z|V_{\mathcal{S}}}^n (\cdot| v_{\mathcal{S}}^n)  \|  Q_{Z}^n )
=\sum_{j=1}^n
D_{\frac{1}{1+s}} ( P_{Z|V_{\mathcal{S}}=v_{\mathcal{S},j}} \|  Q_{Z}   )
=n \sum_{v_{\mathcal{S}}} P_{V_{\mathcal{S}}}(v_{\mathcal{S}})
D_{\frac{1}{1+s}} ( P_{Z|V_{\mathcal{S}}=v_{\mathcal{S}}} \|  Q_{Z}   ),
\end{align}
which follow from the fact that
$v_{\mathcal{S}}^n=(v_{\mathcal{S},1}, \ldots,v_{\mathcal{S},n}) \in 
T_{P_{V_{\mathcal{S}}}}^n$.
Notice that the RHS does not depend on the choice of 
$v_{\mathcal{S}}^n \in T_{P_{V_{\mathcal{S}}}}^n$.
Therefore, we obtain \eqref{NND}.

\begin{remark}
When the conditional distribution $P_{X_i|V_i}$ is deterministic, 
the first inequality in \eqref{XNA} can be simplified to
$ \breve{P}_{X_i^n | V_{i}^n} =  P_{X_i|V_i}^{n}  $.
Therefore, in this case,
\eqref{NND} is simplified as
\begin{align}
s \mathbb{E}_{\cC} \left[
D_{1+s}(  P_{Z^n| \cC}\| P_{Z^n})\right] 
\le 
\sum_{\mathcal{S}\neq \emptyset, \mathcal{S}\subset \{1,2\}}
\frac{ \prod_{i' \in \mathcal{S}} \nu_n(|{\cal X}_{i'}|)
}{ \prod_{i \in \mathcal{S}} \sL_i^s} 
e^{s \breve{I}_{\frac{1}{1-s}}^{\uparrow} ( Z ; V_{\mathcal{S}}) }\Label{NND2}.
\end{align}
\end{remark}

\section{Proof of Lemma \ref{L3C}}\Label{A-G}
\allowdisplaybreaks

We choose 
$Q_{Y^n X_1^n}^*$ as
$Q_{Y^n X_1^n}^*(y^n, x_1^n):=
\frac{\sum_{v_2^n} P_{Y X_1 |V_2}^n(y^n, x_1^n|v_2^n)^{\frac{1}{1+s}}\breve{P}_{V_2^n} (v_2^n)}
{\sum_{y^n, x_1^n,v_2^n} P_{Y X_1 |V_2}^n(y^n, x_1^n|v_2^n)^{\frac{1}{1+s}}\breve{P}_{V_2^n} (v_2^n)
}$.
Then, we have
\begin{align}
e^{-s I_{\frac{1}{1+s}}^{\uparrow} (Y^n X_{1}^n; V_{2}^n)
	[P_{Y X_1 |V_2}^n \times \breve{P}_{V_2^n} ] }
	=  e^{-s D_{\frac{1}{1+s}} ( P_{Y X_1|V_2}^n \times \breve{P}_{V_2^n}  \|  Q_{Y^n X_1^n}^* 
\times \breve{P}_{V_2^n}   )} .
\Label{ZIR}
\end{align}
Since $Q_{Y^n X_1^n}^*$ is permutation invariant,
$Q_{Y^n X_1^n}^*$ is written as a probabilistic mixture of the uniform distribution 
on a fixed-type subset.
Hence, 
there exists a distribution $q$ on $T_n({\cal Y}\times {\cal X}_1)$
such that 
$Q_{Y^n X_1^n}^*=
\sum_{Q_{Y,X_1}\in T_n({\cal Y}\times {\cal X}_1)}
q(Q_{Y,X_1}) \breve{Q}_{Y^n,X_1^n}$.
Then, we have
\begin{align}
&Q_{Y^n X_1^n}^*(y^n x_1^n)^{\frac{s}{1+s}}
=
\Big(\sum_{Q_{Y,X_1}\in T_n({\cal Y}\times {\cal X}_1)}
q(Q_{Y,X_1}) \breve{Q}_{Y^n,X_1^n}(y^n x_1^n)
\Big)^{\frac{s}{1+s}}  \nonumber\\
\le &
\sum_{Q_{Y,X_1}\in T_n({\cal Y}\times {\cal X}_1)}
\Big(
q(Q_{Y,X_1}) \breve{Q}_{Y^n,X_1^n}(y^n x_1^n)
\Big)^{\frac{s}{1+s}}
\le
\sum_{Q_{Y,X_1}\in T_n({\cal Y}\times {\cal X}_1)}
\breve{Q}_{Y^n,X_1^n}(y^n x_1^n)^{\frac{s}{1+s}}
\Label{ZIT}.
\end{align}
Thus, we have
\begin{align}
&e^{-\frac{s}{1+s} D_{\frac{1}{1+s}} ( P_{Y X_1|V_2}^n \times \breve{P}_{V_2^n}  \|  Q_{Y^n X_1^n}^* 
\times \breve{P}_{V_2^n}   )} 
=
\sum_{v_2^n \in T_{P_{V_2}}^n } 
\breve{P}_{V_2^n}(v_2^n)
\sum_{y^n x_1^n}
P_{Y X_1|V_2}^n (y^n x_1^n|v_2^n)^{\frac{1}{1+s}}
Q_{Y^n X_1^n}^*(y^n x_1^n)^{\frac{s}{1+s}}
 \nonumber\\
	\stackrel{(a)}{\le}  &
\sum_{v_2^n \in T_{P_{V_2}}^n } 
\breve{P}_{V_2^n}(v_2^n)
\sum_{y^n x_1^n}
P_{Y X_1|V_2}^n (y^n x_1^n|v_2^n)^{\frac{1}{1+s}}
\sum_{Q_{Y,X_1}\in T_n({\cal Y}\times {\cal X}_1)}
\breve{Q}_{Y^n,X_1^n}(y^n x_1^n)^{\frac{s}{1+s}}  \nonumber\\
=&
\sum_{Q_{Y,X_1}\in T_n({\cal Y}\times {\cal X}_1)}
\sum_{v_2^n \in T_{P_{V_2}}^n } 
\breve{P}_{V_2^n}(v_2^n)
\sum_{y^n x_1^n}
P_{Y X_1|V_2}^n (y^n x_1^n|v_2^n)^{\frac{1}{1+s}}
\breve{Q}_{Y^n,X_1^n}(y^n x_1^n)^{\frac{s}{1+s}}  \nonumber\\
	\stackrel{(b)}{\le}  &
\nu_n(|{\cal Y}| |{\cal X}_1|)^{\frac{s}{1+s}}
\sum_{Q_{Y,X_1}\in T_n({\cal Y}\times {\cal X}_1)}
\sum_{v_2^n \in T_{P_{V_2}}^n } 
\breve{P}_{V_2^n}(v_2^n)
\sum_{y^n x_1^n}
P_{Y X_1|V_2}^n (y^n x_1^n|v_2^n)^{\frac{1}{1+s}}
Q_{Y,X_1}^n(y^n x_1^n)^{\frac{s}{1+s}}  \nonumber\\
=&
\nu_n(|{\cal Y}| |{\cal X}_1|)^{\frac{s}{1+s}}
\sum_{Q_{Y,X_1}\in T_n({\cal Y}\times {\cal X}_1)}
\sum_{v_2^n \in T_{P_{V_2}}^n } 
\breve{P}_{V_2^n}(v_2^n)
e^{-\frac{s}{1+s} 
D_{\frac{1}{1+s}} ( P_{Y X_1|V_2}^n(\cdot |v_2^n) \|  Q_{Y X_1}^n    )} \nonumber\\
\le &
\nu_n(|{\cal Y}| |{\cal X}_1|)^{\frac{s}{1+s}}
|T_n({\cal Y}\times {\cal X}_1)|
\max_{Q_{Y,X_1}}
\sum_{v_2^n \in T_{P_{V_2}}^n } 
\breve{P}_{V_2^n}(v_2^n)
e^{-\frac{s}{1+s} 
D_{\frac{1}{1+s}} ( P_{Y X_1|V_2}^n(\cdot |v_2^n) \|  Q_{Y X_1}^n    )} \nonumber\\
	\stackrel{(c)}{=}  &
	|T_n({\cal Y}\times {\cal X}_1)|
\nu_n(|{\cal Y}| |{\cal X}_1|)^{\frac{s}{1+s}}
\max_{Q_{Y X_1}}
e^{-\frac{s}{1+s} n \sum_{v_2} P_{V_2}(v_2)
D_{\frac{1}{1+s}} ( P_{Y X_1|V_2=v_{2}} \|  Q_{Y X_1}   )} \nonumber\\
	= & 
	|T_n({\cal Y}\times {\cal X}_1)|\nu_n(|{\cal Y}| |{\cal X}_1|)^{\frac{s}{1+s}}
e^{-\frac{s}{1+s} \breve{I}_{\frac{1}{1+s}}^{\uparrow} (Y  X_{1};  V_{2})}  \nonumber\\
	= &
	|T_n({\cal Y}\times {\cal X}_1)|
\nu_n(|{\cal Y}| |{\cal X}_1|)^{\frac{s}{1+s}}
e^{-\frac{s}{1+s} \breve{I}_{\frac{1}{1+s}}^{\uparrow} (Y ;  V_{2}| X_{1})} ,\Label{NBD}
\end{align}
where steps $(a)$ and $(b)$
follow from
\eqref{ZIT} and \eqref{eq3-X}, respectively.
Also, $(c)$ follows from the equations 
\begin{align}
D_{\frac{1}{1+s}} ( P_{Y X_1|V_2}^n(\cdot |v_2^n) \|  Q_{Y X_1}^n    )
=\sum_{j=1}^n
D_{\frac{1}{1+s}} ( P_{Y X_1|V_2=v_{2,j}} \|  Q_{Y X_1}   )
=n \sum_{v_2} P_{V_2}(v_2)
D_{\frac{1}{1+s}} ( P_{Y X_1|V_2=v_{2}} \|  Q_{Y X_1}   ),
\end{align}
which follow from the fact that
$v_2^n=(v_{2,1}, \ldots,v_{2,n}) \in T_{P_{V_2}}^n$.

In the same way as \eqref{MLD},
we have
\begin{align}
	P_{Y|X_1 X_2}^n \cdot \breve{P}_{X_2^n|V_2^n}\times \breve{P}_{X_1^n} 
	\le &  \nu_n(|{\cal X}_1|)\nu_n(|{\cal X}_2|) 
	P_{Y|X_1 X_2}^n \cdot P_{X_2|V_2}^n\times P_{X_1}^n 
	\nonumber \\
	=&  \nu_n(|{\cal X}_1|)\nu_n(|{\cal X}_2|) 
	P_{Y X_1 |V_2}^n \Label{MZR}.
\end{align}

Therefore, we have
\begin{align}
	\mathbb{E}_{\cC'} \left[ e_{\cC'}\right]
	\stackrel{(a)}{\le}  &
     (\sN-1)^s
	e^{-s I_{\frac{1}{1+s}}^{\uparrow} (Y^n X_{1}^n; V_{2}^n)
	[P_{Y|X_1 X_2}^n \cdot \breve{P}_{X_2^n|V_2^n}\times \breve{P}_{X_1^n}\times \breve{P}_{V_2^n} ] } \nonumber\\
	\stackrel{(b)}{\le} & (\sN-1)^s \nu_n(|{\cal X}_1|)\nu_n(|{\cal X}_2|) e^{-s I_{\frac{1}{1+s}}^{\uparrow} (Y^n X_{1}^n; V_{2}^n)
	[P_{Y X_1 |V_2}^n \times \breve{P}_{V_2^n} ] } \nonumber\\
	\stackrel{(c)}{\le}  &
	|T_n({\cal Y}\times {\cal X}_1)|^{1+s} \nu_n(|{\cal Y}| |{\cal X}_1| )^{s} 
	\nu_n(|{\cal X}_1|)\nu_n(|{\cal X}_2|) 
\sN^s 
e^{-s \breve{I}_{\frac{1}{1+s}}^{\uparrow} (Y ;  V_{2}| X_{1})} ,
	\end{align}
where steps $(a)$, $(b)$, and $(c)$ follow from 
\eqref{NMD} of Lemma \ref{L3}, 
the combination of \eqref{MVK} and \eqref{MZR}, and 
the combination of \eqref{ZIR} and \eqref{NBD}, 
respectively.
Therefore, we obtain \eqref{ZMH}.

\begin{remark}
When the conditional distribution $P_{X_i|V_i}$ is deterministic, 
\eqref{MZR} is simplified as
\begin{align}
	P_{Y|X_1 X_2}^n \cdot \breve{P}_{X_2^n|V_2^n}\times \breve{P}_{X_1^n} 
	\le & 
	 \nu_n(|{\cal X}_1|)
	P_{Y X_1 |V_2}^n .
\end{align}
Thus,
\eqref{ZMH} is simplified as
\begin{align}
\mathbb{E}_{\cC'} \left[
e_{\cC'}\right]
\le 
	|T_n({\cal Y}\times {\cal X}_1)|^{1+s} \nu_n(|{\cal Y}| |{\cal X}_1| )^{s} 
	\nu_n(|{\cal X}_1|)
\sN^s
e^{-s \breve{I}_{\frac{1}{1+s}}^{\uparrow} (Y ;  V_{2}| X_{1})}.
\end{align}
\end{remark}

\section{Proof of Theorem \ref{add-F}}\label{sec: proof of Theorem add-F}
Since 
$H(Y_2|V_1, V_2)\ge H(Y_2|V_1, V_2, X_1, X_2)=H(Y_2|X_1, X_2)= H(N_2)$, we have
\begin{align}
	I(Y_2;V_1|V_2)-I(Z;V_1)
	\le
	I(Y_2;V_1|V_2)
	\le \log q- H(N_2).
\end{align}
In the same way, we have
\begin{align}
	I(Y_1;V_2|V_1)-I(Z;V_2)
	\le \log q- H(N_1).
\end{align}

We assume that
there exists an independent random variable $N_1'$ from $N_1$ 
such that 
$Z=a_3a_1^{-1} Y_1+N_1'$.
Then, we have
\begin{align}
	&I(Y_2;V_1|V_2) + I(Y_1;V_2|V_1) - I(V_1, V_2;Z) \nonumber\\
	=&
	I(Y_2;V_1|V_2) + I(V_1, V_2; Y_1) -I(Y_1;V_1)
	- I(V_1, V_2;Z)  \nonumber\\
	\stackrel{(a)}{=}&
	I(Y_2;V_1|V_2) + I(Y_1;V_1, V_2|Z) -I(Y_1;V_1)  \nonumber\\
	\stackrel{(b)}{\le} &
	I(Y_2;V_1|V_2) + I(Y_1;X_1, X_2|Z) -I(Y_1;V_1)  \nonumber\\
	=&
	H(Y_2|V_2)- H(Y_2|V_1, V_2) + I(Y_1;X_1, X_2|Z) -I(Y_1;V_1)  \nonumber\\
	\stackrel{(c)}{\le} &
	H(Y_2|V_2)- H(Y_2|X_1, X_2) + I(Y_1;X_1, X_2|Z) -I(Y_1;V_1) 
	\nonumber
	\\
	\stackrel{(d)}{=} &
	H(Y_2|V_2)- H(Y_2|X_1, X_2) + 
	I(Y_1;X_1, X_2) 
	-I(Z;X_1, X_2) 
	-I(Y_1;V_1)  \nonumber
	\\
	= &
	H(Y_2|V_2)- H(Y_2|X_1, X_2) + 
	H(Y_1)-H(Z)
	-H(Y_1|X_1, X_2) 
	+H(Z|X_1, X_2) 
	-I(Y_1;V_1)  \nonumber
	\\
	\stackrel{(e)}{\le} &
	\log q - H(Y_2|X_1, X_2)  
	-H(Y_1|X_1, X_2) 
	+H(Z|X_1, X_2) 
	\nonumber \\
	=&\log q - H(N_2)+H(N_3) - H(N_1),
\end{align}
where  $(a)$, $(b)$ and $(d)$ follows by the Markov chain
$(V_1,V_2)-(X_1,X_2)-Y_1-Z$ which holds since $Z=a_3a_1^{-1} Y_1+N_1';$ $(c)$ follows by the Markov chain
$(V_1,V_2)-(X_1,X_2)-Y_2;$ and $(e)$ is due to the fact that $H(Y_1)\le H(Z)$ as $Z=a_3a_1^{-1} Y_1+N_1'.$

Therefore, all conditions in 
${\cal R}_{J,N}$ and ${\cal R}_{I,N}$
are upper bounded by the conditions in the set in the RHS of 
\eqref{HH3} and \eqref{HH4}.
Hence,
the equalities in \eqref{HH3} and \eqref{HH4} hold.

\section{Proof of Theorem \ref{add-C}}\label{sec: proof of Theorem add-C}
To prove Theorem \ref{add-C}, we need to show the validity of \eqref{HH4-7B} and \eqref{HH4-6}, the proofs of which are given as follows.
\subsection{Proof of \eqref{HH4-7B}}
First we show that $h(q)-H(N_2)$ serves as an outer bound on $R_1$
even with adaptive codes with the 1st type decoder.
Fano's inequality implies
\begin{align}
	nR_1- I(M_1;M_2, X_2^n) 
	=& H(M_1 )- I(M_1;M_2, X_2^n) \nonumber\\
	=&
	H(M_1|M_2, X_2^n ) \nonumber\\ 
	=& H(M_1|M_2, X_2^n,Y_2^n) +I(M_1;Y_2^n|M_2, X_2^n) \nonumber \\
	\le & 
	\log 2 +\epsilon \log M_1 
	+I(M_1;Y_2^n|M_2, X_2^n).
\end{align}
Thus, we have
\begin{align}
	nR_1
	&\le \log 2 +\epsilon \log M_1 +I(M_1;Y_2^n|M_2,X_2^n)+ I(M_1;M_2,X_2^n)  \nonumber\\
	&= \log 2 +\epsilon \log M_1 +I(M_1;M_2,X_2^n, Y_2^n).
	\Label{HH0}
\end{align}

Consider
\begin{align}
	I(M_1;M_2,X_2^n, Y_2^n)=&
	I(M_1;X_2^n, Y_2^n|M_2)  \nonumber\\
	=& I(M_1;Y_2^n|M_2) + I(M_1;X_2^n|M_2, Y_2^n) \nonumber \\
	\stackrel{(a)}{=} & I(M_1;Y_2^n|M_2)   \nonumber\\
	=& \sum_{j=1}^n I(M_1; Y_{2,j}|M_2,Y_{2}^{j-1})  \nonumber \\
	\stackrel{(b)}{\le}  &  \sum_{j=1}^n H(Y_{2,j})-H(Y_{2,j}|M_1, M_2, X_{1,j}, X_{2,j}, Y_{2}^{j-1})  \nonumber \\
	\stackrel{(c)}{=}  &
	\sum_{j=1}^n H(Y_{2,j})-H(Y_{2,j}|X_{1,j},X_{2,j})  \nonumber\\
	\le & n (\log q- H(N_2))\Label{HH11},
\end{align}
where 
$(a)$ follows by the Markov chain $M_1 - (M_2,Y_{2}^{n}) - X_{2}^n$, which is shown by an inductive use of the Markov chain $M_1 - (M_2,X_{2}^{j-1},Y_{2}^{j-1}) - X_{2,j};$
$(b)$ follows by the fact that conditioning reduces entropy; and $(c)$ follows from the Markov chain $(M_1, M_2,X_1^{j-1}, X_2^{j-1},Y_1^{j-1},Y_2^{j-1}, Z^{j-1})- 
(X_{1,j},X_{2,j})-(Y_{1,j}, Y_{2,j}, Z_j)$.
Combining \eqref{HH0} and \eqref{HH11}, we have
\begin{align}
	n R_1 \le n (\log q- H(N_2)) +\log 2 +\epsilon \log M_1,
\end{align}
which implies $R_1 \le \log q- H(N_2)$.

A similar proof can be applied to show that $\log q-H(N_1)$ serves as an outer bound on $R_2.$

\subsection{Proof of \eqref{HH4-6}}
Consider the sum-rate $R_1+R_2$ for non-adaptive codes with the 1st type decoder under joint secrecy constraint. We have
\begin{align}
	n(R_1+R_2)	
	=& H(M_1|M_2) + H(M_2|M_1)\nonumber \\
	=& H(M_1| X_2^n, Y_2^n,M_2) + I(M_1; X_2^n, Y_2^n|M_2)
	+H(M_2| X_1^n, Y_1^n,M_1) + I(M_2; X_1^n, Y_1^n|M_1)\nonumber \\
	=& H(M_1| X_2^n, Y_2^n,M_2) 
	+ I(M_1; Y_2^n| X_2^n,M_2)
	+ I(M_1;  X_2^n|M_2)\nonumber \\
	&+H(M_2| X_1^n, Y_1^n,M_1) 
	+ I(M_2; Y_1^n| X_1^n,M_1)
	+ I(M_2; X_1^n|M_1)\nonumber \\
	\stackrel{(a)}{=}& H(M_1|X_2^n,  Y_2^n,M_2) 
	+ I(M_1; Y_2^n| X_2^n,M_2)
	+H(M_2| X_1^n, Y_1^n,M_1) + I(M_2;Y_1^n| X_1^n,M_1)\nonumber \\
	\stackrel{(b)}{\leq}& I(M_1;Y_2^n|X_2^n,M_2)
	+I(M_2;Y_1^n|X_1^n,M_1)+2n\mathcal{O}(\epsilon_n)\nonumber \\
	\stackrel{(c)}{=}& I(M_1;Y_2^n|X_2^n)
	+I(M_2;Y_1^n|X_1^n)+2n\mathcal{O}(\epsilon_n)\nonumber \\
	\stackrel{(d)}{\leq}& I(M_1;Y_2^n|X_2^n)+I(M_2;Y_1^n|X_1^n)-I(M_1,M_2;Z^n)+2n\mathcal{O}(\epsilon_n)+n\tau_n, \Label{NBA}
\end{align}
where $(a)$ from the fact that for non-adaptive codes, 
$M_1$ and $X_1^n$ are independent of $M_2$ and $X_2^n;$ $(b)$ follows from Fano's inequality;
$(c)$ follows from the Markovian chains
$M_2-X_2^n-(M_1,Y_2^n) $ and
$M_1-X_1^n-(M_2,Y_1^n) $;
$(d)$ follows from the joint secrecy constraint as defined in \eqref{eq:SSec}.

Since $N_3=
\frac{b_3}{b_1} {N}_1+  \frac{a_3}{a_2}{N}_2+N_3'$,
we have $a_3 {X}_1^n+b_3 {X}_2^n+{N}_3^n=
b_3 {X}_2^n+\frac{b_3}{b_1} {N}_1^n+
(a_3 {X}_1^n+\frac{a_3}{a_2}{N}_2^n+N_3')
$. Hence,
we have $H(a_3 {X}_1^n+b_3 {X}_2^n+{N}_3^n)
\ge H(b_3 {X}_2^n+\frac{b_3}{b_1} {N}_1^n)$.
Thus, 
\begin{align}
	H\Big( a_3 {X}_1^n+\frac{a_3}{a_2}{N}_2^n\Big)
	+H\Big( b_3 {X}_2^n+\frac{b_3}{b_1} {N}_1^n\Big)
	-H(a_3 {X}_1^n+b_3 {X}_2^n+{N}_3^n)
	\le n \log q.
\end{align}
Thus,
using Lemma \ref{MN9} in Appendix \ref{App:Lemma}, we have
\begin{align}
	& I(M_1;Y_2^n|X_2^n)+I(M_2;Y_1^n|X_1^n)-I(M_1,M_2;Z^n)\nonumber \\
	\le &
	H\Big( a_3 {X}_1^n+\frac{a_3}{a_2}{N}_2^n\Big)
	-H\Big(\frac{a_3}{a_2} {N}_2^n\Big)
	+H\Big( b_3 {X}_2^n+\frac{b_3}{b_1} {N}_1^n\Big)
	-H\Big(\frac{b_3}{b_1} {N}_1^n\Big)
	-H(a_3 {X}_1^n+b_3 {X}_2^n+{N}_3^n)
	+H({N}_3^n)  \nonumber \\
	\le &n (\log q +H(N_3) - H(N_1)- H(N_2)).\Label{NAS}
\end{align}
The combination of \eqref{NBA} and \eqref{NAS} implies
\begin{align}
	R_1+R_2 \le \log q +H(N_3) - H(N_1)- H(N_2).
\end{align}

\section{Useful lemmas for additive channels}\Label{App:Lemma}
First we prepare a useful lemma for additive channels.
Assume that ${\cal X}_1$, ${\cal X}_2$, ${\cal Y}_1$, ${\cal Y}_2$,
and ${\cal Z}$ are over a general field $\mathbb{K}$, which is not necessarily a finite field.
We consider the additive channels given in \eqref{AD1} -- \eqref{AD3} or \eqref{AG1} -- \eqref{AG3} over the field $\mathbb{K}$, where all coefficients are elements of $\mathbb{K}$.
Then, we have the following lemma.


\begin{lemma}\Label{MN9}
When 
there exists 
independent random variable $N_3'$ from $N_1$ and $N_2$
such that $N_3=
\frac{b_3}{b_1} {N}_1+  \frac{a_3}{a_2}{N}_2+N_3'$,
we have
\begin{align*}
	& I(M_1; Y_2^n|X_2^n)+ I(M_2; Y_1^n|X_1^n) - I(M_1, M_2; Z^n)\\
\le 
	&
	H\Big( a_3 X_1^n+\frac{a_3}{a_2}N_2^n\Big)-H\Big(\frac{a_3}{a_2} N_2^n\Big)+H\Big( b_3 X_2^n+\frac{b_3}{b_1} N_1^n\Big)-H\Big(\frac{b_3}{b_1} N_1^n\Big)
	-H(a_3 X_1^n+b_3 X_2^n+N_3^n)+H(N_3^n)  .
\end{align*}
\end{lemma}

\begin{proofof}{Lemma \ref{MN9}}
We have
\begin{align*}
	& I(M_1; Y_2^n|X_2^n)+ I(M_2; Y_1^n|X_1^n) - I(M_1, M_2; Z^n)\\
	= 
	& 
	H(a_2 X_1^n+ N_2^n)-H(a_2 X_1^n+ N_2^n|M_1)+ H(b_1 X_2^n+ N_1^n)-H(b_1 X_2^n+ N_1^n|M_2)
	\\
	&-(H(a_3 X_1^n+ b_2 X_2^n+ N_3^n)-H(a_3 X_1^n+ b_3 X_2^n+ N_3^n|M_1, M_2)  )\\
	\stackrel{(a)}{\le} 
	&
	H(a_2 X_1^n+ N_2^n)-H(a_2 X_1^n+ N_2^n|M_1 M_2)+H(b_1 X_2^n+ N_1^n)-H(b_1 X_2^n+ N_1^n|M_1, M_2)
	\\
	&-(  H(a_3 X_1^n+ b_3 X_2^n+ N_3^n)-H(a_3 X_1^n+ b_3 X_2^n+ N_3^n|M_1, M_2)  )\\
	=&
	H\Big( a_3 X_1^n+\frac{a_3}{a_2}N_2^n\Big)-H\Big(\frac{a_3}{a_2}N_2^n\Big)+H\Big( b_3 X_2^n+\frac{b_3}{b_1} N_1^n\Big)-H\Big(\frac{b_3}{b_1} N_1^n\Big)
	-H(a_3 X_1^n+b_3 X_2^n+N_3^n)+H(N_3^n)  \\
	&-\Big(H\Big( a_3 X_1^n+\frac{a_3}{a_2}N_2^n\Big|M_1,  M_2\Big)-H\Big(\frac{a_3}{a_2}N_2^n\Big)+H\Big( b_3 X_2^n+\frac{b_3}{b_1} N_1^n\Big|M_1,  M_2\Big)-H\Big(\frac{b_3}{b_1} N_1^n\Big)\\
	&-H(a_3 X_1^n+b_3 X_2^n+N_3^n|M_1, M_2) +H(N_3^n)\Big) \\
	\stackrel{(b)}{\le} 
	&
	H\Big( a_3 X_1^n+\frac{a_3}{a_2}N_2^n\Big)-H\Big(\frac{a_3}{a_2} N_2^n\Big)+H\Big( b_3 X_2^n+\frac{b_3}{b_1} N_1^n\Big)-H\Big(\frac{b_3}{b_1} N_1^n\Big)
	-H(a_3 X_1^n+b_3 X_2^n+N_3^n)+H(N_3^n)  ,
\end{align*}
where
$(a)$ follows from the fact that conditioning reduces entropy; 
$(b)$ follows from the proof below.
\begin{align*}
	&H\Big( a_3 X_1^n+\frac{a_3}{a_2}N_2^n\Big|M_1,  M_2\Big)-H\Big(\frac{a_3}{a_2}N_2^n\Big)+H\Big( b_3 X_2^n+\frac{b_3}{b_1} N_1^n\Big|M_1,  M_2\Big)-H\Big(\frac{b_3}{b_1} N_1^n\Big)\\
	&-H(a_3 X_1^n+b_3 X_2^n+N_3^n|M_1, M_2) +H(N_3^n)\\
	=&
	I\Big( a_3 X_1^n+\frac{a_3}{a_2}N_2^n;a_3 X_1^n\Big|M_1,  M_2\Big)+I\Big(b_3 X_2^n+\frac{b_3}{b_1} N_1^n;b_3 X_2^n \Big|M_1, M_2 \Big)\\
	&-I\Big(a_3 X_1^n+b_3 X_2^n+N_3^n; a_3 X_1^n+b_3 X_2^n\Big|M_1, M_2\Big)\\
	\stackrel{(c)}{=} &
	I\Big(a_3 X_1^n+\frac{a_3}{a_2}N_2^n, b_3 X_2^n+\frac{b_3}{b_1} N_1^n;a_3 X_1^n, b_3 X_2^n \Big|M_1, M_2\Big)
	-I\Big(a_3 X_1^n+b_3 X_2^n+N_3^n; a_3 X_1^n+b_3 X_2^n\Big|M_1, M_2\Big) \\
	\stackrel{(d)}{\ge} &
	I\Big(a_3 X_1^n+b_3 X_2^n+\frac{b_3}{b_1} N_1^n+  \frac{a_3}{a_2}N_2^n; a_3 X_1^n+b_3 X_2^n\Big|M_1, M_2\Big)
	-I\Big(a_3 X_1^n+b_3 X_2^n+N_3^n; a_3 X_1^n+b_3 X_2^n\Big|M_1, M_2\Big)\\ 
	\stackrel{(e)}{\ge}& 0,
\end{align*}
where $(c)$ holds because $X_1^n$ and $X_2^n$ are independent and they are also independent of $N_1^n, N_2^n$; $(d)$ follows from the non-negativity of the mutual information; and $(e)$ follows from the assumption 
$N_3=
\frac{b_3}{b_1} {N}_1+  \frac{a_3}{a_2}{N}_2+N_3'$.
\end{proofof}

Moreover, for the additive Gaussian TWC as defined in \eqref{AG1} -- \eqref{AG3}. We have the following lemma:
\begin{lemma}\Label{L4}
	Assume the condition
	\begin{align}
		\mathbb{E}[\|X_{1}^n\|^2]\leq  n c_1 ,\quad
		\mathbb{E}[\|X_{2}^n\|^2]\leq  n c_2
		\Label{Con1}.
	\end{align}
	For non-adaptive codes, when $ \frac{b_3^2}{b_1^2}v_1+ \frac{a_3^2}{a_2^2}v_2\le v_3$, we have 
	\begin{align}
		& H\Big( a_3 X_1^n+\frac{a_3}{a_2}N_2^n\Big)+H\Big( b_3 X_2^n +\frac{b_3}{b_1} N_1^n\Big)
		-H(a_3 X_1^n+b_3 X_2^n+N_3^n)  \nonumber\\
		\le & \frac{n}{2} (\log 2\pi e (b_3^2 c_2+\frac{b_3^2}{b_1^2}v_1)
		+\log 2\pi e (a_3^2 c_1+\frac{a_3^2}{a_2^2}v_2)
		-\log 2\pi e (a_3^2 c_1+ b_3^2 c_2+v_3)).
	\end{align}
\end{lemma}

\begin{proofof}{Lemma \ref{L4}}
We define the functions
\begin{align}
	g(x):=& \frac{n}{2}\log (2\pi e \frac{x}{n}) \Label{def: func g(x)}\\
	h(y,z):=& g(z+g^{-1}(y))-y, 
\end{align}
where $g^{-1}$ is the inverse function of $g$.
Hence, we have
\begin{align}
	g^{-1}(y)= \frac{n e^{2\frac{y}{n}}}{2\pi e} .
\end{align}
Under the average power constraints, the fact that $N_1^n, N_2^n$ are independent of $X_1^n, X_2^n$, and the maximum differential entropy lemma, we have
\begin{align}
	H( b_3 X_2^n+\frac{b_3}{b_1} N_1^n) &\le g(n b_3^2 c_2+\frac{b_3^2}{b_1^2}nv_1)\\
	H( a_3 X_1^n+\frac{a_3}{a_2} N_2^n)&\le g(na_3^2 c_1+\frac{a_3^2}{a_2^2}nv_2).
\end{align}

As a simple extension of \cite[Lemma 8]{Leung-Yan}, we have the following lemma.
\begin{lemma}
	\Label{L5}
	For a fixed $z$, $h(y,z)$ is monotonically decreasing for $y$.
\end{lemma}

\begin{lemma}[\protect{\cite[Eq. (5)]{Blachman}}]
	\Label{L6}
	When $X_1^n$ and $X_2^n$ are independent $n$-dimensional vectors
	\begin{align}
		e^{\frac{2}{n}H(X_1^n+X_2^n)} \ge e^{\frac{2}{n} H(X_1^n)} + e^{\frac{2}{n} H(X_2^n)} .
	\end{align}
	The equality holds when $X_1^n$ and $X_2^n$ are Gaussian random variables with proportional covariance matrices.
\end{lemma}

We define a Gaussian random variable $N_3'$ with mean 0 and variance
$v_3':=v_3-\frac{b_3^2}{b_1^2}v_1- \frac{a_3^2}{a_2^2}v_2$ that is independent of other random variables
$X_1,X_2,N_1,N_2,N_3$. 
Since $X_1^n$ and $X_2^n$ are independent of each other, using Lemma \ref{L6}, 
we have
\begin{align}
	H(a_3 X_1^n+b_3 X_2^n+N_3) 
	=& H(a_3 X_1^n+b_3 X_2^n+\frac{b_3}{b_1} N_1^n+\frac{a_3}{a_2}N_2^n+ N_3') \nonumber\\
	\ge & \frac{n}{2}\log (
	e^{\frac{2}{n}H(b_3 X_2^n+\frac{b_3}{b_1} N_1^n)}
	+e^{\frac{2}{n}H(a_3 X_1^n+\frac{a_3}{a_2}N_2^n+ N_3')})\nonumber\\
	\ge & \frac{1}{2}\log (
	e^{\frac{2}{n}H( b_3 X_2^n+\frac{b_3}{b_1} N_1^n)}
	+e^{\frac{2}{n}H(a_3 X_1^n+\frac{a_3}{a_2}N_2^n)}
	+e^{\frac{2}{n}H(N_3')}) \nonumber\\
	=& \frac{n}{2}\log (
	e^{\frac{2}{n}H(b_3 X_2^n+\frac{b_3}{b_1} N_1^n)}
	+e^{\frac{2}{n}H(a_3 X_1^n+\frac{a_3}{a_2}N_2^n)}
	+e^{\frac{2}{n} g (n v_3')}).\Label{EE1}
\end{align}
Hence, 
we have
\begin{align}
	& H(a_3 X_1^n+b_3 X_2^n+N_3^n) -
	H( b_3 X_2^n+\frac{b_3}{b_1} N_1^n)-H(a_3 X_1^n+\frac{a_3}{a_2}N_2^n) \nonumber\\
	\stackrel{(a)}{\ge} &
	\frac{n}{2}\log (
	e^{\frac{2}{n}H(b_3 X_2^n+\frac{b_3}{b_1} N_1^n)}
	+e^{\frac{2}{n}H(a_3 X_1^n+\frac{a_3}{a_2}N_2^n)}
	+e^{\frac{2}{n}g (n v_3')})
	-H(b_3 X_2^n+\frac{b_3}{b_1} N_1^n)-H(a_3 X_1^n+\frac{a_3}{a_2}N_2^n) \nonumber\\
	=&
	\frac{n}{2}\log \frac{2}{n}\pi e(
	g^{-1}(H(b_3 X_2^n+\frac{b_3}{b_1} N_1^n))
	+g^{-1} (H(a_3 X_1^n+\frac{a_3}{a_2}N_2^n))
	+g^{-1}(g (n v_3')))  \nonumber\\
	&-H(b_3 X_2^n+\frac{b_3}{b_1} N_1^n)-H(a_3 X_1^n+\frac{a_3}{a_2}N_2^n) \nonumber\\
	=&
	g(
	g^{-1}(H(b_3 X_2^n+\frac{b_3}{b_1} N_1^n))
	+g^{-1} (H(a_3 X_1^n+\frac{a_3}{a_2}N_2^n))
	+n v_3') \nonumber\\
	&
	-H(b_3 X_2^n+\frac{b_3}{b_1} N_1^n)-H(a_3 X_1^n+\frac{a_3}{a_2}N_2^n)
	\nonumber\\
	\stackrel{(b)}{\ge} &
	g(
	g^{-1}(g(n b_3^2 c_2+\frac{b_3^2}{b_1^2}n v_1))
	+g^{-1} (H(a_3 X_1^n+\frac{a_3}{a_2}N_2^n))
	+n v_3')
	-g(n b_3^2 c_2+\frac{b_3^2}{b_1^2} n v_1)-H(a_3 X_1^n+\frac{a_3}{a_2}N_2^n) \nonumber\\
	\stackrel{(c)}{\ge} &
	g(g^{-1}(g(n b_3^2 c_2+\frac{b_3^2}{b_1^2}n v_1))
	+g^{-1} (g(n a_3^2 c_1+\frac{a_3^2}{a_2^2}n v_2))
	+n v_3')
	-g(n b_3^2 c_2+\frac{b_3^2}{b_1^2}n v_1)- g(n a_3^2 c_1+\frac{a_3^2}{a_2^2}n v_2) \nonumber\\
	= &
	g(n a_3^2 c_1+\frac{a_3^2}{a_2^2}n v_2
	+n b_3^2 c_2+\frac{b_3^2}{b_1^2}n v_1
	+n v_3')
	-g(n b_3^2 c_2+\frac{b_3^2}{b_1^2}n v_1)- g(n a_3^2 c_1+\frac{a_3^2}{a_2^2}n v_2) \nonumber\\
	\stackrel{(d)}{=} &
	g(n a_3^2 c_1
	+n b_3^2 c_2+n v_3)
	-g(n b_3^2 c_2+\frac{b_3^2}{b_1^2}n v_1)- g(n a_3^2 c_1+\frac{a_3^2}{a_2^2}n v_2) \nonumber\\
	\stackrel{(e)}{=} &  \frac{n}{2} \Big(\log 2\pi e (a_3^2 c_1+ b_3^2 c_2+v_3))-\log 2\pi e (b_3^2 c_2+\frac{b_3^2}{b_1^2}v_1)
	-\log 2\pi e (a_3^2 c_1+\frac{a_3^2}{a_2^2}v_2)\Big),	
\end{align}
where each step is shown as follows.
$(a)$ follows from \eqref{EE1}.
$(b)$ follows from Lemma \ref{L5} with $y= H(b_3 X_2^n+\frac{b_3}{b_1} N_1^n)$ and $z=g^{-1} (H(a_3 X_1^n+\frac{a_3}{a_2}N_2^n))
+n v_3'$.
$(c)$ follows from Lemma \ref{L5} with $y= H(a_3 X_1^n+\frac{a_3}{a_2}N_2^n)$ and 
$z=g^{-1}g((n a_3^2 c_1+\frac{a_3^2}{a_2^2}n v_2)+n v_3'));$ 
$(d)$ follows the fact that $v_3=\frac{b_3^2}{b_1^2}v_1+ \frac{a_3^2}{a_2^2}v_2+v_3';$ and $(e)$ follows from the definition of function $g(\cdot)$ in \eqref{def: func g(x)}. Therefore, we obtain the desired statement. 
\end{proofof}

\section{Proof of Theorem \ref{G-C}}\label{sec: proof of Theorem G-C}
The outer bounds for $R_1$ and $R_2$ are already implied by the capacity region for Gaussion TWC as given in \cite{Han1984}. 
Here we only need to show the outer bound on the sum-rate $R_1+R_2$. 

In the same way as \eqref{NBA}, we have
\begin{align}
	n(R_1+R_2)	
	\le I(M_1;Y_2^n|X_2^n)+I(M_2;Y_1^n|X_1^n)-I(M_1,M_2;Z^n)+2n\mathcal{O}(\epsilon_n)+n\tau_n. \Label{NBA2}
\end{align}

Since it is sufficient to show the inequality
\begin{align}
	R_1+R_2 \le&
	\frac{1}{2} \Big(
	\log (b_1^2 c_2+v_1)
	+\log (a_2^2 c_1+v_2)
	-\log (a_3^2 c_1+ b_3^2 c_2+v_3)
	+\log v_3 -\log v_1- \log v_2 \Big),	\Label{AAI}
\end{align}
for non-adaptive codes under the condition $ \frac{b_3^2}{b_1^2}v_1+ \frac{a_3^2}{a_2^2}v_2\le v_3,$
the desired statement follows from 
the following lemma.

\begin{lemma}\Label{L7}
	Assume the condition \eqref{Con1}. 
	For non-adaptive codes, when $ \frac{b_3^2}{b_1^2}v_1+ \frac{a_3^2}{a_2^2}v_2\le v_3$, we have
	\begin{align*}
		\frac{1}{n} \Big(I(M_1; &Y_2^n|X_2^n)+I(M_2; Y_1^n|X_1^n)- I( M_1,  M_2; Z^n)\Big)\\
		\le &
		\frac{1}{2} \Big(
		\log (b_1^2 c_2+v_1)
		+\log (a_2^2 c_1+v_2)
		-\log (a_3^2 c_1+ b_3^2 c_2+v_3) 
		+\log  v_3 -\log  v_1- \log  v_2) \Big).
	\end{align*}
\end{lemma}

\begin{proofof}{Lemma \ref{L7}}
In order to prove Lemma \ref{L7},
we apply Lemma \ref{MN9} to Gaussian two-way wiretap channel. Then, we have
\begin{align}
	& I(M_1; Y_2^n|X_2^n)+ I(M_2; Y_1^n|X_1^n) - I(M_1, M_2; Z^n) \nonumber\\
	\stackrel{(a)}{\le} &
	H\Big( a_3 X_1^n+\frac{a_3}{a_2}N_2^n\Big)-H\Big(\frac{a_3}{a_2} N_2^n\Big)+H\Big( b_3 X_2^n+\frac{b_3}{b_1} N_1^n\Big)-H\Big(\frac{b_3}{b_1} N_1^n\Big)
	-H(a_3 X_1^n+b_3 X_2^n+N_3^n)+H(N_3^n)   \nonumber\\
	\stackrel{(b)}{\le} 
	&\frac{n}{2} \Big(
	\log 2\pi  e(b_3^2 c_2+\frac{b_3^2}{b_1^2}v_1)
	+\log 2\pi e (a_3^2 c_1+\frac{a_3^2}{a_2^2}v_2)
	-\log 2\pi  e(a_3^2 c_1+ b_E^2 c_2+v_3)  \nonumber\\
	&+\log 2\pi e v_3 -\log 2\pi e \frac{b_3^2}{b_1^2} v_1- \log 2\pi e \frac{a_3^2}{a_2^2} v_2) \Big), \nonumber\\
	=& \frac{n}{2} \Big(
	\log 2\pi  e(b_1^2 c_2+v_1)
	+\log 2\pi e (a_2^2 c_1+v_2)
	-\log 2\pi  e(a_3^2 c_1+ b_3^2 c_2+v_3)  \nonumber\\
	&+\log 2\pi e v_3 -\log 2\pi e v_1- \log 2\pi e v_2) \Big),
\end{align}
where
$(a)$ follows from Lemma \ref{MN9};
$(b)$ follows from Lemma \ref{L4}. 
Then, we obtain Lemma \ref{L7}.
\end{proofof}

\end{document}